\documentclass[12pt,a4paper]{article}
\usepackage{amsmath,amssymb,amsfonts,amsthm}
\newtheorem{theorem}{Theorem}[section]
\newtheorem{lemma}[theorem]{Lemma}
\newtheorem{definition}[theorem]{Definition}

\usepackage{graphicx}

\newcommand{\Su}{\mathcal{S}}

\newcommand{\mf}{\mathcal{M}}

\newcommand{\mfs}{\mathcal{M}^{\Su}}

\newcommand{\mq}{m_{bh}}

\newcommand{\ds}{ds}

\newcommand{\dv}{dv}

\newcommand{\Rt}{\mathbb{R}^3}

\newcommand{\Ls}{\Delta_0}

\newcommand{\kif}{\bar\eta}

\title{Geometric inequalities for axially symmetric black holes}

\author{Sergio Dain\\
\\
 Facultad de Matem\'atica, Astronom\'\i a y F\'\i sica, FaMAF, \\
Universidad Nacional de C\'ordoba \\
Instituto de F\'\i sica Enrique Gaviola, IFEG, CONICET, \\
Ciudad Universitaria (5000) C\'ordoba, Argentina.\\
\\
Max-Planck-Institut f{\"u}r Gravitationsphysik, Albert Einstein\\
Institut, Am M\"uhlenberg 1 D-14476 Potsdam Germany}

\begin{document}
\maketitle

\begin{abstract} 
  A geometric inequality in General Relativity relates quantities that have
  both a physical interpretation and a geometrical definition. It is well known
  that the parameters that characterize the Kerr-Newman black hole satisfy
  several important geometric inequalities. Remarkably enough, some of these
  inequalities also hold for dynamical black holes. This kind of inequalities
  play an important role in the characterization of the gravitational collapse,
  they are closed related with the cosmic censorship conjecture.  Axially
  symmetric black holes are the natural candidates to study these inequalities
  because the quasi-local angular momentum is well defined for them. We review
  recent results in this subject and we also describe the main ideas behind the
  proofs. Finally, a list of relevant open problem is presented.
\end{abstract}

\section{Introduction}
\label{sec:introduction}


Geometric inequalities have an ancient history in Mathematics. A classical
example is the isoperimetric inequality for closed plane curves given by
\begin{equation}
  \label{eq:54}
  L^2 \geq 4\pi A,
\end{equation}
where $A$ is the area enclosed by a curve $C$ of length $L$, and where equality
holds if and only if $C$ is a circle (for a  review on this subject see
\cite{Osserman78}). General Relativity is a geometric theory, hence it is not
surprising that geometric inequalities appear naturally in it.  As we will see,
many of these inequalities are similar in spirit as the isoperimetric
inequality (\ref{eq:54}). However, General Relativity as a physical theory
provides an important extra ingredient. It is often the case that the
quantities involved have a clear physical interpretation and the expected
behavior of the gravitational and matter fields often suggest geometric
inequalities which can be highly non-trivial from the mathematical point of
view.  The interplay between geometry and physics gives to geometric
inequalities in General Relativity their distinguished character.

A prominent example is the positive mass theorem. The physics suggests that the
mass of the spacetime (which is represented by a pure geometrical quantity
\cite{Arnowitt62}\cite{Bartnik86}\cite{chrusciel86}) should be positive and
equal to zero if and only if the spacetime is flat.  From the geometrical mass
definition, without the physical picture, it would be very hard to conjecture
this inequality. In fact the proof turn out to be very subtle
\cite{Schoen79b}\cite{Schoen81}\cite{witten81}.

A key assumption in the positive mass theorem is that the matter fields should
satisfy an energy condition.  This condition is expected to hold for all
physically realistic matter. It is remarkable that such a simple condition
encompass a huge class of physical models and that it translates into a pure
geometrical condition.  This kind of general properties which do not depend
very much on the details of the model are not easy to find for astrophysical
objects (like stars or galaxies) which usually have a very complicated
structure. And hence it is difficult to obtain simple geometric inequalities
among the parameters that characterize them. 

In contrast, black holes represent a unique class of very simple macroscopic
objects that play, in some sense, the role of `elementary particles' in the
theory. The black hole uniqueness theorem ensures that stationary black holes
in electro-vacuum are characterized by three parameters, which can be taken to
be the area $A$ of the black hole, the angular momentum $J$ and the charge
$q$. The mass $m$ is calculated in terms of these parameters by an explicit
formula (cf. equation (\ref{eq:11})).  It is well known that these parameters
satisfy certain geometrical inequalities which restrict the range of them.
These inequalities are direct consequences of the explicit formula
(\ref{eq:11}).  Among them, we note first the following
\begin{equation}
  \label{eq:2bc}
  m\geq \sqrt{\frac{A}{16\pi}},
\end{equation}
which will lead to the Penrose inequality for dynamical black holes. Also we
have the following two inequalities which will play a central role in this
article
\begin{equation}
  \label{eq:24i}
  m^2\geq \frac{q^2+\sqrt{q^4+4J^2}}{2}, \quad  A \geq  4\pi \sqrt{q^4+4J^2} .
\end{equation}
The equality in (\ref{eq:2bc}) is achieved for the Schwarzschild black hole. The
equality in both inequalities \eqref{eq:24i} are achieved for extreme black
holes.

However black holes are not stationary in general. Astrophysical phenomena like
the formation of a black hole by gravitational collapse or a binary black hole
collision are highly dynamical. For such systems, the black hole can not be
characterized by few parameters as in the stationary case. In fact, even
stationary but non-vacuum black holes have a complicated structure (for example
black holes surrounded by a rotating ring of matter, see the numerical studies
in \cite{Ansorg05}). Remarkably, inequalities (\ref{eq:2bc})--(\ref{eq:24i})
extend (under appropriate assumptions) to the fully dynamical regime. Moreover,
inequalities (\ref{eq:2bc})--(\ref{eq:24i}) are deeply connected with
properties of the global evolution of Einstein equations, in particular with
the cosmic censorship conjecture. The main subject of this review is to present
a series of recent results which are mainly concerned with the dynamical
versions of inequalities (\ref{eq:24i}).

To extend the validity of inequalities (\ref{eq:2bc})--(\ref{eq:24i}) to
non-stationary black holes the first difficulty is how to define the physical
parameters involved, most notably the angular momentum $J$ of a dynamical black
hole.  To define quasi-local quantities is in general a difficult problem (see
the review \cite{Szabados04}). However, for axially symmetric black holes, the
angular momentum (via Komar's formula) is well defined and it is conserved in
vacuum.  Essentially for this reason inequalities (\ref{eq:24i}) have been
mostly studied for axially symmetric black holes. An exception are inequalities
which involves only the electric charge since the charge is well defined as a
quasi-local quantity without any symmetry assumption.

The plan of the article is the following. In section \ref{sec:physical-picture}
we describe the heuristic physical arguments that support these inequalities
and connect them with global properties of a gravitational collapse. In section
\ref{sec:results} we present an overview of the main results concerning these
inequalities that have been recently obtained. We also describe the main ideas
behind the proofs. Two important geometrical quantities involved in
(\ref{eq:24i}), mass and angular momentum, have distinguished properties in
axial symmetry (in contrast to the electric charge).  These properties play a
fundamental role. We describe in some detail the angular momentum in section
\ref{sec:angul-moment-axial} and the mass in section
\ref{sec:mass-axial-symmetry}. Finally in section \ref{sec:open-problems} we
present the relevant open problem in this area.

\section{The physical picture}
\label{sec:physical-picture}
The most important example of a geometric inequality for dynamical black holes
is the Penrose inequality.  In a seminal article Penrose \cite{Penrose73}
proposed a physical argument that connects global properties of the
gravitational collapse with geometric inequalities on the initial
conditions. For a recent review about this inequality see \cite{Mars:2009cj}
and references therein.  Since it will play an important role in what follows,
let us review Penrose argument.

We will assume that the following statements hold in a gravitational collapse:
\begin{itemize}

\item[(i)]  Gravitational collapse results in a black hole (weak cosmic
censorship). 

\item[(ii)]  The spacetime settles down to a stationary final
state. We will further assume that at some finite time all the matter have
fallen into the black hole and hence the exterior region is
electro-vacuum. 
\end{itemize}
Conjectures (i) and (ii) constitute the standard picture of the gravitational
collapse. Relevant examples where this picture is confirmed (and where the role
of angular momentum is analyzed) are the collapse of neutron stars studied
numerically in \cite{Baiotti:2004wn} \cite{Giacomazzo:2011cv}.

Before going into the Penrose argument, let us analyze the final stationary
state postulated in (ii). The black hole uniqueness theorem implies that the
final state is given by the Kerr-Newman black hole (we emphasize however that
many important aspects of the black holes uniqueness still remain open, see
\cite{Chrusciel:2008js} for a recent review on this problem).  Let us denote by
$m_0$, $A_0$, $J_0$ and $q_0$ the mass, area, angular momentum and charge of
the remainder Kerr-Newman black hole.  In order to describe a black hole, the
parameters of the Kerr-Newman family of solutions of Einstein equations should
satisfy the remarkably inequality
\begin{equation}
  \label{eq:13}
  d\geq 0, 
\end{equation}
where we have defined 
\begin{equation}
  \label{eq:8}
  d= m_0^2-q_0^2-\frac{J_0^2}{m_0^2}.
\end{equation}
This inequality is equivalent to 
\begin{equation}
  \label{eq:24}
  m_0^2\geq \frac{q_0^2+\sqrt{q_0^4+4J_0^2}}{2}.
\end{equation} 
From Newtonian considerations (take $q=0$ for simplicity), we can interpret
this inequality as follows (see \cite{Wald71}): in a collapse the gravitational
attraction ($\approx m_0^2/r^2$) at the horizon ($r\approx m_0$) dominates over the
centrifugal repulsive forces ( $\approx J_0^2/m_0r^3$).
It is important to recall that the Kerr-Newman solution is well defined for any
choice of the parameters, but it only represents a black hole if the inequality
(\ref{eq:24}) is satisfied. 

The Kerr-Newman black hole is called extreme if the equality in (\ref{eq:24})
is satisfied, namely
\begin{equation}
  \label{eq:16}
  m_0^2 =  \frac{q_0^2+\sqrt{q_0^4+4J_0^2}}{2}.
\end{equation}
The area of the black hole horizon is given by the important formula
\begin{equation}
   \label{eq:11}
  A_0=4\pi \left(2m^2_0-q_0^2+2m_0 \sqrt{d}  \right).
\end{equation}
Note that this expression has meaning only if the inequality (\ref{eq:24})
holds.  

\begin{figure}
  \centering
   \includegraphics[width=0.5\textwidth]{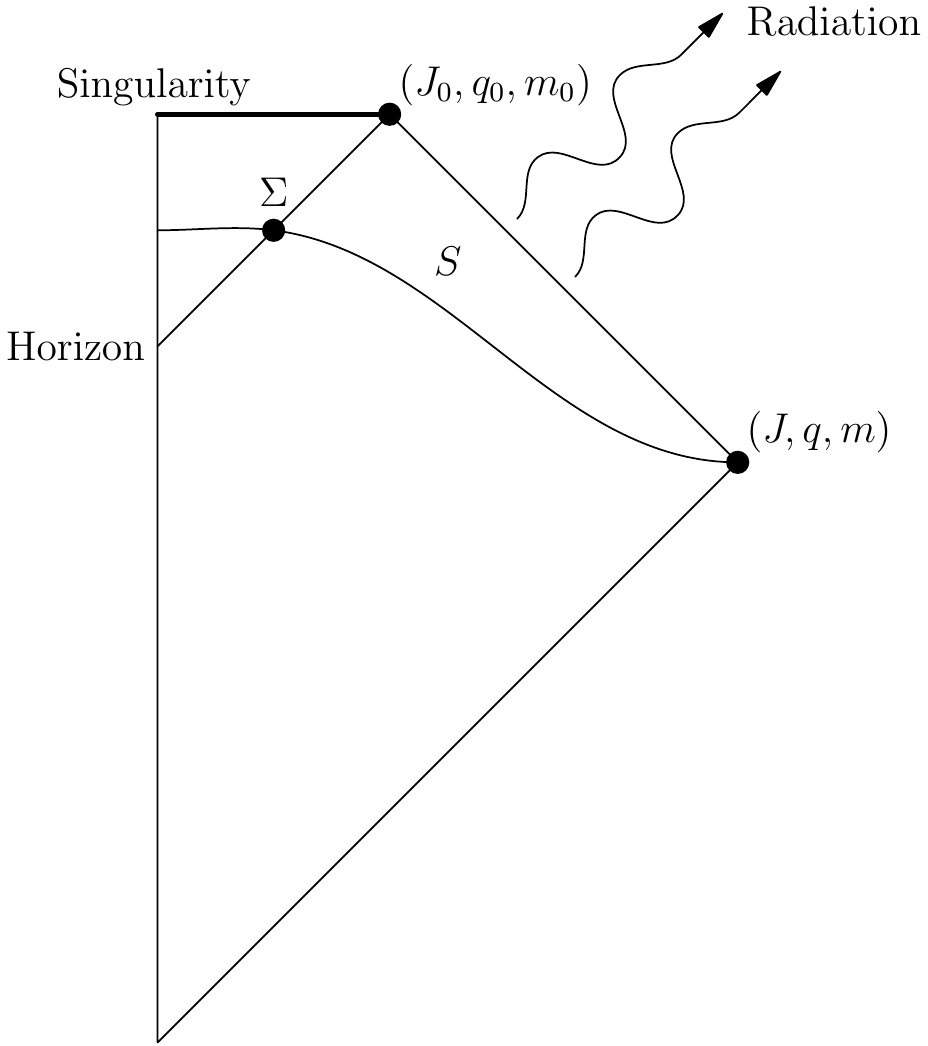} 
  \caption{Schematic representation of a gravitational collapse.}
  \label{fig:1}
\end{figure}

Penrose argument runs as follows.  Let us consider a gravitational
collapse. Take a Cauchy surface $S$ in the spacetime such that the collapse has
already occurred. This is shown in figure \ref{fig:1}.
Let $\Sigma$ denotes the intersection of the event horizon with the Cauchy
surface $S$ and let $A$ be its area.  Let $(m,q,J)$ be the total mass, charge
and angular momentum at spacelike infinity. These quantities can be computed
from the initial surface $S$. By the black hole area theorem we have that the
area of the black hole increase with time and hence
\begin{equation}
  \label{eq:15}
  A_0\geq A.
\end{equation}
Since gravitational waves carry positive energy, the total mass of the
spacetime should be bigger than the final mass of the black hole
\begin{equation}
  \label{eq:4}
  m\geq m_0.
\end{equation}
The difference $m-m_0$ is the total amount of gravitational radiation emitted
by the system.

The area $A_0$ of the remainder black hole is given by equation (\ref{eq:11})
in terms of the final parameters ($J_0, q_0, m_0$). It is a monotonically
increasing function of $m_0$ (for fixed $q_0$ and $J_0$), namely the derivative
$\partial A_0/\partial m_0$ is positive (we explicitly calculate this
derivative bellow). Then, using this monotonicity and inequalities
(\ref{eq:15}) and (\ref{eq:4}) we  obtain
\begin{equation}
  \label{eq:19c}
  A\leq A_0 \leq 4\pi \left(2m^2-q_0^2+2m
    \left(m^2-q_0^2-\frac{J_0^2}{m^2}\right)^{1/2}  \right),
\end{equation}
where the important point is that in the right hand side appears the total mass
$m$ (instead of $m_0$), which can be calculated on the Cauchy surface $S$. The
parameters $q_0$ and $J_0$ are not known a priori but since they appear with
negative sign we have
\begin{equation}
  \label{eq:20c}
 A\leq A_0 \leq  4\pi \left(2m^2-q_0^2+2m
    \left(m^2-q_0^2-\frac{J_0^2}{m^2}\right)^{1/2}  \right)\leq 16\pi m^2.
\end{equation}
Remains still an important point: how to estimate the area $A$ of $\Sigma$ in
terms of geometrical quantities that can be locally computed on the initial
conditions. Recall that in order to know the location of the event horizon the
whole spacetime is needed.  Assume that the surface $S$ contains a future
trapped two-surface $\Sigma_0$ (the trapped condition is a local property). By a
general result on black hole spacetimes we have that the surface $\Sigma_0$
should be contained in $\Sigma$. But that does not necessarily means that the
area of $\Sigma_0$ is smaller than the area of $\Sigma$.  Consider all surfaces
$\tilde \Sigma$ enclosing $\Sigma_0$. Denote by $A_{\min}(\Sigma_0)$ the
infimum of the areas of all such surfaces. Then we clearly have that
$A(\Sigma)\geq A_{\min}(\Sigma_0)$.  The advantage of this construction is that
$A_{\min}(\Sigma_0)$ is a quantity that can be computed from the Cauchy surface
$S$. Using this inequality and inequalities (\ref{eq:19c}) and (\ref{eq:20c})
we finally obtain the Penrose inequality
\begin{equation}
  \label{eq:58}
  m\geq \sqrt{\frac{ A_{\min}(\Sigma_0)}{16\pi}}.
\end{equation}
For further discussion we refer to \cite{Mars:2009cj} and references therein.

Penrose argument is remarkable because its end up in an inequality that can be
written purely in terms of the initial conditions. On the other hand, the proof
of such inequality gives indirect evidences of the validity of the conjectures
(i) and (ii).

Can we include the parameters $q$ and $J$ in the inequality
(\ref{eq:58}) to get a stronger version of it? The problem is, of course, how
to relate the final state parameters $(q_0, J_0)$ with the initial state ones
$(q,J)$.
   
If the matter fields are not charged, then the charge is conserved, namely 
\begin{equation}
  \label{eq:21}
  q=q_0.
\end{equation}
And hence in that case we have the following version of the Penrose inequality
with charge
\begin{equation}
  \label{eq:22}
  A_{\min}(\Sigma_0)\leq  A\leq  4\pi \left(2m^2-q^2+2m
    \left(m^2-q^2\right)^{1/2}  \right). 
\end{equation}

The case of angular momentum is more complicated. Angular momentum is in
general non-conserved. There exists no simple relation between the total
angular momentum $J$ of the initial conditions and the angular momentum $J_0$
of the final black hole. For example, a system can have $J=0$ initially, but
collapse to a black hole with final angular momentum $J_0\neq 0$. We can
imagine that on the initial conditions there are two parts with opposite
angular momentum, one of them falls in to the black hole and the other scape to
infinity.

Axially symmetric vacuum spacetimes constitute a remarkable exception because
the angular momentum is conserved. In that case we have
\begin{equation}
  \label{eq:59}
  J=J_0.
\end{equation}
We discuss this conservation law in detail in section
\ref{sec:axially-symm-init}.  The physical interpretation of (\ref{eq:59}) is
that axially symmetric gravitational waves do not carry angular momentum.

For non-vacuum axially symmetric spacetimes the angular momentum is no longer
conserved. Matter can transfer angular momentum even in axial symmetry.  This
is also true for the electromagnetic field. However in the electro-vacuum case
a remarkably effect occurs. First, the sum of the gravitational and the
electromagnetic angular momentum is conserved. Second, at spacelike infinity
only the gravitational angular momentum is non-zero.  In section
\ref{sec:angul-moment-axial} we prove this two facts. Hence if $J$ and $J_0$
denotes now the total angular momentum, then the conservation (\ref{eq:59})
still holds for axially symmetric electro-vacuum spacetimes and we obtain the
full Penrose inequality valid for axially symmetric electro-vacuum initial
conditions
\begin{equation}
  \label{eq:23}
A_{\min}(\Sigma_0)\leq  A\leq 4\pi \left(2m^2-q^2+2m
    \left(m^2-q^2-\frac{J^2}{m^2}\right)^{1/2}  \right). 
\end{equation}
We emphasize that in this inequality the total angular momentum $J$ can be
computed at any closed surface that surround the black hole using the formula
(\ref{eqcm:21}).  When this surface
is at infinity the angular momentum is given by the gravitational angular
momentum (i.e. the Komar integral of the axial Killing field).

Inequality (\ref{eq:23}) implies the bound
\begin{equation}
  \label{eq:60}
 m^2\geq \frac{q^2+\sqrt{q^4+4J^2}}{2}.
\end{equation}
Of course, inequality (\ref{eq:60}) can be deduced directly with the same
argument without using the area theorem.  The first place where this conjecture
was formulated  is in \cite{Friedman82} (see also \cite{Horowitz84}). 

Inequality (\ref{eq:60}) can be viewed as a simplified version of the Penrose
inequality. The major difference is that the area of the horizon does not
appears in (\ref{eq:60}). Only charges, which are essentially topological,
appear in the right hand side of this inequality.


Inequality (\ref{eq:60}) is a global inequality for two reasons. First, it
involves the total mass $m$ of the spacetime. Second it assumes global
restrictions on the initial data: axial symmetry and electro-vacuum. We will
discuss these assumptions in more detail in section \ref{sec:results}.

The area $A$, the angular momentum $J$ in axial symmetry, and the charge $q$
are quasi-local quantities (in particular, the right hand side of (\ref{eq:60})
is purely quasi-local). Namely they carry information on a bounded region of
the spacetime. In contrast with a local quantity like a tensor field which
depends on a point of the spacetime or a global quantities (like the total
mass) which depends on the whole initial conditions.  A natural question is
whether dynamical black holes satisfy purely quasi-local inequalities. The
relevance of this kind of inequalities is that they provide a much finer
control on the dynamics of a black holes than the global versions.

It is well known that the energy of the gravitational field can not be
represented by a local quantity. The best one can hope is to obtain a
quasi-local expression.  These are the so called quasi-local mass definition
(see the review article \cite{Szabados04} and reference therein).  Consider the
formula (\ref{eq:11}) for the horizon area for the Kerr-Newman black
holes. From this expression, we can write the mass in terms of the other
parameters as follows
\begin{equation}
  \label{eq:32}
  \mq= \sqrt{\frac{A}{16\pi}+ \frac{q^2}{2}+\frac{\pi(q^4 +4J^2)}{A}}.
\end{equation}
In this equation we have dropped the subindice $0$ in the right hand side and
also denoted the mass in the left hand side by $\mq$ to emphasize that this
expression can be in principle defined for any black hole (i.e. not necessarily
stationary). With this interpretation, this expression is known as the
Christodoulou \cite{Christodoulou70} mass of the black hole.
 
For a dynamical black hole the expression (\ref{eq:32}) is in principle just a
definition.  Does the formula \eqref{eq:32} represents the quasi-local mass of
a non-stationary black hole? Let us analyze its physical behavior. We discuss
first the relation of this quasi-local mass with the total mass of the
spacetime.  For only one black hole we expect the following inequality to
be true
\begin{equation}
  \label{eq:53}
  m \geq \mq.
\end{equation}
This inequality implies the Penrose inequality (\ref{eq:23}) but it is stronger
(see the discussion in \cite{Mars:2009cj}).  However, it is important to
emphasize that for the case of many black holes this inequality does not holds. In
fact it is possible to find counter examples if we take the area as additive
\cite{Weinstein05} or the quasi-local masses as additives
\cite{Dain:2010pj}. This is expected, since the interaction energy of the black
holes need to be taken into account (see the discussion in \cite{Dain:2010pj}).
For only one black hole, the inequality (\ref{eq:53}) in axial symmetry is an
open problem in this general form, we will further discuss it in section
\ref{sec:open-problems}.

We discuss now the purely quasi-local properties of (\ref{eq:32}). 
The formula \eqref{eq:32} trivially satisfies the inequality (\ref{eq:60}).
This is, of course, just because the Kerr black hole satisfies this bound.
Hence, if we accept \eqref{eq:32} as the correct formula for the quasi-local
mass of an axially symmetric black hole, then (\ref{eq:32}) provides the,
rather trivial, quasi-local version of \eqref{eq:60}.  

Consider the evolution of $\mq$. By the area theorem, we know that the horizon
area will increase. If we assume axial symmetry and electro-vacuum, then the
total angular momentum (gravitational plus electromagnetic) will be conserved
at the quasi-local level. On physical grounds, one would expect that in this
situation the quasi-local mass of the black hole should increase with the area,
since there is no mechanism at the classical level to extract mass from the
black hole.  In effect, the only way to extract mass from a black hole is by
extracting angular momentum through a Penrose process.  But angular momentum
transfer is forbidden in electro-vacuum axial symmetry.  Then, one would expect
that both the area $A$ and the quasi-local mass $\mq$ should monotonically
increase with time.

Let us take a time derivative of $\mq$ (denoted by a dot).  To analyze this, it
is illustrative to write down the complete differential, namely the first law of
thermodynamics
\begin{equation}
  \label{eq:55}
  \delta \mq=  \frac{\kappa}{8 \pi} \delta A + \Omega_H \delta J + \Phi_H
  \delta q,
\end{equation}
where
\begin{equation}
  \label{eq:56}
  \kappa= \frac{1}{4 \mq} \left(1- \left(\frac{4\pi}{A}\right)^2 (q^4+4J^2)\right),  
\quad \Omega_H= \frac{4\pi J}{A\mq}, \quad  \Phi_H= \frac{4\pi (\mq+\sqrt{d})
  q}{A}, 
\end{equation}
where $\mq$ is given by (\ref{eq:32}) and $d$ (defined in equation
(\ref{eq:8})) is written in terms of $A$ and $J$ and $q$ as
\begin{equation}
  \label{eq:17}
  d= \frac{1}{\mq^2}\left(\frac{A}{16\pi}\right)^2 \left(1-(q^4+4J^2)\left(\frac{4\pi}{A} \right)^2 \right)^2.
\end{equation}
In equation (\ref{eq:55}) we have followed the standard notation for the
formulation of the first law, we emphasize however that in our context this
equation is a trivial consequence of (\ref{eq:32}).

Under our assumptions, from the formula
\eqref{eq:32} we obtain
\begin{equation}
  \label{eq:34}
  \dot m_{bh} = \frac{\kappa }{8\pi}\dot A , 
\end{equation}
were we have used that the angular momentum $J$ and the charge $q$ are
conserved. Since, by the area theorem, we have
\begin{equation}
  \label{eq:9}
  \dot A \geq 0,
\end{equation}
the time derivative of $\mq$ will be positive (and hence the mass $\mq$ will
increase with the area) if and only if $\kappa \geq 0$, that is 
\begin{equation}
  \label{eq:5}
  4\pi \sqrt{q^4+4J^2} \leq  A.
\end{equation}
Then, it is natural to conjecture that \eqref{eq:5} should be satisfied for any
black hole in an axially symmetry.  If the horizon violate
\eqref{eq:5} then in the evolution the area will increase but the mass $\mq$
will decrease. This will indicate that the quantity $\mq$ has not the desired
physical meaning.  Also, a rigidity statement is expected. Namely, the equality
in \eqref{eq:5} is reached only by the extreme Kerr black hole given by the
formula
\begin{equation}
  \label{eq:6}
  A = 4\pi\left (\sqrt{q^4+4J^2 }\right). 
\end{equation}
The final picture is that the size of the black hole is bounded from bellow by
the charge and angular momentum, and the minimal size is realized by the
extreme Kerr-Newman black hole.  This inequality provides a remarkable
quasi-local measure of how far a dynamical black hole is from the extreme case,
namely an `extremality criteria' in the spirit of \cite{Booth:2007wu}, although
restricted only to axial symmetry.  In the article \cite{Dain:2007pk} it has
been conjectured that, within axially symmetry, to prove the stability of a
nearly extreme black hole is perhaps simpler than a Schwarzschild black
hole. It is possible that this quasi-local extremality criteria will have
relevant applications in this context.  Note also that the inequality
\eqref{eq:5} allows to define, at least formally, the positive surface gravity
density (or temperature) of a dynamical black hole by the formula (\ref{eq:56})
(see Refs. \cite{Ashtekar03} \cite{Ashtekar02} for a related discussion of the
first law in dynamical horizons).

If inequality \eqref{eq:5} is true, then we have a non trivial monotonic
quantity (in addition to the black hole area) $\mq$ in electro-vacuum
\begin{equation}
  \label{eq:10b}
 \dot m_{bh} \geq 0.
\end{equation}

It is important to emphasize that the physical arguments presented above in
support of \eqref{eq:5} are certainly weaker in comparison with the ones behind
the Penrose inequalities (\ref{eq:58}), (\ref{eq:22}) and (\ref{eq:23}).  A
counter example of any of these inequality will prove that the standard picture
of the gravitational collapse is wrong. On the other hand, a counter example of
\eqref{eq:5} will just prove that the quasi-local mass \eqref{eq:32} is not
appropriate to describe the evolution of a non-stationary black hole.  One can
imagine other expressions for quasi-local mass, may be more involved, in axial
symmetry.  On the contrary, reversing the argument, a proof of \eqref{eq:5}
will certainly suggest that the mass \eqref{eq:32} has physical meaning for
non-stationary black holes as a natural quasi-local mass (at least in axial
symmetry). Also, the inequality \eqref{eq:5} provide a non trivial control of
the size of a black hole valid at any time.


Finally, it is important to explore the physical scope of validity of these
geometrical inequalities. Are they valid for other macroscopic objects?  The
Penrose inequality (\ref{eq:23}) is clearly not valid for an arbitrary region
in the spacetime.  Namely, consider an arbitrary 2-surface of area $A$, which
is not necessarily a black hole boundary. We can make $A$ arbitrary large
keeping the mass (total or quasi-local) small (for example, take a region in
Minkowski).

For the inequalities (\ref{eq:60}) and (\ref{eq:5}) the situation is less
obvious.  To have an intuitive idea of the order of magnitude involved it is
important to include the relevant constants in these inequalities.  Let $G$ be
the gravitational constant and $c$ the speed of light. Then, these inequalities
are written as follows
\begin{equation}
  \label{eq:4b}
   m^2\geq \frac{1}{G}\frac{q^2+\sqrt{q^4+4J^2c^2}}{2}.
\end{equation}
and
\begin{equation}
  \label{eq:7b}
   4\pi \frac{G}{c^4}\sqrt{q^4+4 J^2 c^2} \leq  A.
\end{equation}
For the reader's convenience we include the explicit values of the constants
(in centimeters, grams and seconds) 
\begin{equation}
  \label{eq:10c}
  G  =6.67 \times 10^{-8} \, cm^3 g^{-1} s^{-2},\quad 
c  = 3 \times 10^{10} \,cm\, s^{-1}.
\end{equation}
The values of the fundamental physical constants used in this section were
taken from \cite{mohr08}.

Let us analyze first the global inequality (\ref{eq:4b}). It is useful to split
it into the cases with  zero charge and zero angular momentum
respectively, namely
\begin{equation}
  \label{eq:8b}
 \sqrt{G} \geq \frac{|q|}{m},
\end{equation}
and
\begin{equation}
  \label{eq:14}
 \frac{G}{c}\geq \frac{|J|}{m^2}. 
\end{equation}
For the inequality (\ref{eq:8b}) consider an electron and a proton, for these
particles the quotient on the right hand side is given by 
\begin{align}
  \label{eq:qme}
  \frac{|q_e|}{m_e} &=0.53 \times 10^{18} \, g^{-1/2} cm^{3/2}s^{-1},\\
  \frac{|q_e|}{m_p}  & =2.87 \times 10^{14} \, g^{-1/2} cm^{3/2}s^{-1}. \label{eq:qmp}
\end{align}
Since in these units we have $\sqrt{G} \approx 10^{-4}$, we see these particles
grossly violate the global inequality (\ref{eq:8b}). And hence ordinary charged
matter would violate it also (a similar discussion has been presented in
\cite{Gibbons82} and \cite{Horowitz84}).

For the angular momentum case (\ref{eq:14}) we can also consider an elementary
particle. In that case the angular momentum of a particle with spin $s$ (recall
that $s=1/2$ for the electron and the proton), is given by
\begin{equation}
  \label{eq:13b}
  J=\sqrt{s(s+1)}\hbar, \quad   \hbar  =1.05 \times 10^{-27} \, cm^2  s^{-1} g,
\end{equation}
where $\hbar$ is the Planck constant. For example, for the electron we have
\begin{equation}
  \label{eq:17c}
  \frac{|J|}{m^2_e}=1.12 \times 10^{27}\, cm^2  s^{-1} g^{-1}.
\end{equation}
Since 
\begin{equation}
  \label{eq:19d}
  \frac{G}{c} =2.22 \times 10^{-18} \, cm^2  s^{-1} g^{-1}, 
\end{equation}
inequality (\ref{eq:14}) is also violated by several order of magnitude for
elementary particles. Instead of an elementary particle we can consider an
ordinary rotating object. It is clear that there exists ordinary object for
which $J/m^2 \approx 1$ (say a rigid sphere of mass $1 g$, radius $1 cm$, and
angular velocity $1 s^{-1}$) and hence inequality (\ref{eq:14}) is also
violated for ordinary rotating bodies.  

We conclude that ordinary matter does not satisfies in general the global
inequality (\ref{eq:4b}).  This inequality should be interpreted as a property
of electro-vacuum gravitational fields on complete regular initial conditions
where both the charge and the angular momentum are ``produced by the topology''
and not by matter sources (unless, of course, that they are inside a black hole
horizon). By ``produced by the topology'' we mean the following. The angular
momentum and the electric charge are defined as integral over closed two
dimensional surfaces. In electro-vacuum these integrals are conserved (we
discuss this in detail in section \ref{sec:angul-moment-axial}) and hence they
are zero if the topology of the initial conditions is trivial (i.e. $\Rt$). In
order to have non-trivial charges in electro-vacuum the initial conditions
should have some ``holes''.  This non-trivial topology signals the presence of
a black hole.

For the quasi-local inequality (\ref{eq:7b}) it is also convenient to
distinguish between the cases with zero charge  and zero angular momentum 
respectively
\begin{equation}
  \label{eq:9b}
  A\geq \frac{G}{c^4}4\pi q^2,
\end{equation}
and
\begin{equation}
  \label{eq:15b}
  A\geq 8\pi \frac{G}{c^3} |J|.
\end{equation}
Since the charge is discrete, in
unit of $q_e$, it make sense to calculate the following characteristic radius 
\begin{equation}
  \label{eq:20}
  r_0=  \frac{q_e \,G^{1/2}}{c^2}= 1.38 \times 10^{-34}\, cm.
\end{equation}
We see that $r_0$ is one order of magnitude less than the Planck length
$l_p$ given by  
\begin{equation}
  \label{eq:1}
 l_p  =  \left(\frac{G\hbar}{c^3}\right)^{1/2}= 1.6 \times 10^{-33} \,cm.
\end{equation}

If we assume that the particle or the macroscopic object has spherical shape we
can define  the area radius  $r$ by $A=4\pi r^2$. Then, the inequality
(\ref{eq:9b}) for a particle of charge $q_e$ has the form
\begin{equation}
  \label{eq:21b}
  r\geq r_0. 
\end{equation}
The proton charge radius is $r_p = 0.87 \times 10^{-12} \, cm$ according to
\cite{pohl10} and $r_p = 0.84 \times 10^{-12} \, cm$ according to the recent
calculation presented in \cite{mohr08}.  Hence, the proton satisfies inequality
(\ref{eq:21b}). This inequality is also consistent with the upper bound for the
electron radius $10^{-20} \, cm$ measured in \cite{dehmelt88}.

For the case of angular momentum, using the relation (\ref{eq:13b}) we can
compute the following quotient for an elementary particle
\begin{equation}
  \label{eq:22b}
  r_0=\left( 2\frac{G}{c^3}|J|\right)^{1/2}=\sqrt{2} (s(s+1))^{1/4} l_p.
\end{equation}
We see, that for a particle of spin $s$ of order $1$ the minimal radius is of
the order of the Planck length $l_p$ and hence it is also satisfied for
elementary particles.

More relevant is the case of an ordinary rotating
body. The angular momentum $J$ of a rigid body is given by
\begin{equation}
  \label{eq:64m}
  J=I \omega,
\end{equation}
where $I$ is the moment of inertia and $\omega$ the angular velocity. 
Consider an ellipsoid of revolution  with semi-axes $a$
and $b$, rotating along the $b$ axis. The moment of inertia along the axis of
rotation is given by
\begin{equation}
  \label{eq:65}
  I=\frac{2}{5}m a^2,
\end{equation}
where $m$ is the mass of the ellipsoid, which is assumed to have constant
density. The area of the ellipsoid satisfies the following elementary
inequality
\begin{equation}
  \label{eq:87}
  A \geq 2\pi a^2.
\end{equation}
The equality in (\ref{eq:87}) is achieved in the limit $b\to 0$, namely, the
bound is sharp.

Inserting equations (\ref{eq:64m}) and (\ref{eq:65}) in to the inequality
(\ref{eq:15b}), and using the bound (\ref{eq:87}) and the fact that it is sharp,
we obtain that the inequality (\ref{eq:15b}) is satisfied by the ellipsoid if
and only if the following inequality holds
\begin{equation}
  \label{eq:50}
  \frac{5}{8}\frac{c^3}{G}\geq m \omega.
\end{equation}
It is interesting to note that in the inequality (\ref{eq:50}) neither the area
nor the radii of the body appear.  The inequality relates only the mass and the
angular velocity of the body. Also, the body is not assumed be nearly
spherical, the parameters for the ellipsoid are arbitrary.

The value of the left hand side of this inequality is
\begin{equation}
  \label{eq:57}
\frac{5}{8}\frac{c^3}{G}= 2.53\times  10^{38}\, s^{-1} g.
\end{equation}
 For the sun we have the following values
\begin{equation}
  \label{eq:66h}
  m_{sun}=1.989 \times 10^{33}\, g,\quad 
 \quad  \omega= 2.90 \times 10^{-6} rad\, s^{-1},
\end{equation}
and hence
\begin{equation}
  \label{eq:52}
  m \omega =5.77 \times10^{27} \, s^{-1} g.
\end{equation}
We see that the inequality is satisfied for the sun.  In order to violate
(\ref{eq:50}) a body should be very massive and highly spinning, a natural
candidate for that is a neutron star. For the fastest spinning neutron star
found to date (see \cite{Hessels:2006ze}) we have
\begin{equation}
  \label{eq:66}
  \omega \approx 4.5 \times 10^3 \, rad\, s^{-1}.
\end{equation}
Assuming that the neutron stars has about three solar masses (which appears to
be a reasonable upper bound for the mass, see \cite{Lattimer:2004pg}) we obtain
\begin{equation}
  \label{eq:67}
  m \omega\approx 2.7 \times 10^{37}\, s^{-1} g.
\end{equation}
The inequality (\ref{eq:50}) is still satisfied however the value (\ref{eq:67})
is remarkable close to the upper limit  (\ref{eq:57}).

The example of the ellipsoid shows the elementary relation between shape and
angular momentum in classical mechanics which is valid even for non spherical
bodies.  There is no such relation between shape and the electric charge of a
body. In fact, we will present some counter examples for charged highly
prolated objects that violate the inequality (\ref{eq:9b}).  However,
remarkably enough, for charged `round surfaces' (we will define this concept
later on), an inequality between area and charge can be proved (see theorem
\ref{t:main3}).  On the other hand, the example of the ellipsoid
suggests that the scope of validity of the inequality between area and angular
momentum (\ref{eq:15b}) (or the related inequality (\ref{eq:50})) for axial
symmetric bodies is much larger.

\section{Results and main ideas}
\label{sec:results}
In this section we present the main results concerning inequalities
(\ref{eq:60}) and (\ref{eq:5}) that haven been recently proved in the
literature. We also discuss the general strategy of their proofs. 

\subsection{Global inequality}
\label{sec:global-inequality}


The proof of the inequality between total mass and charge (namely, setting
$J=0$ in (\ref{eq:60})), which is valid without any symmetry assumptions, has
been known for some time. The first proof was provided in \cite{Gibbons82} and
\cite{Gibbons83} using spinorial arguments similar to the Witten proof of the
positive mass theorem \cite{witten81}.  See also \cite{Horowitz84}. A related
inequality was proved in \cite{Moreschi84} with similar techniques. In
\cite{Bartnik05} the proof was generalized to include low differentiable
metrics. For this inequality an interesting rigidity result is expected: the
equality holds if and only if the initial data are embed into the
Majumdar-Papapetrou static spacetime (see \cite{Hartle:1972ya} for a discussion
of these spacetimes in relations with black holes). The rigidity statement has
also been established, but with supplementary hypotheses, in \cite{Gibbons82}
and \cite{Chrusciel:2005ve}.  Very recently a new proof was provided in
\cite{Khuri11} which removes all remaining hypothesis. Also in this article a
new approach is presented. The strategy is to combine the Jang equation method
with the spinorial proof of the positive mass theorem.


The inclusion of angular momentum in axial symmetry (which is the main subject
of this review) involves complete different techniques. In particular no
spinorial proof of these inequalities are available so far (see however
\cite{Zhang99} where a related inequality is proved using spinors).  The first
proof of the global inequality (\ref{eq:60}) (with no electric charge) was
provided in a series of articles \cite{Dain05c}, \cite{Dain05d}, \cite{Dain05e}
which end up in the global proof given in \cite{Dain06c}.

In \cite{Chrusciel:2007dd} and \cite{Chrusciel:2007ak} the result was
generalized and the proof simplified. In \cite{Chrusciel:2009ki}
\cite{Costa:2009hn} the charge was included. As a sample of the most general
result currently available we present the following theorem proved in
\cite{Chrusciel:2009ki} and 
\cite{Costa:2009hn}.
 
\begin{theorem}
\label{t:main-1}
Consider an axially symmetric, electro-vacuum, asymptotically flat and maximal
initial data set with two asymptotics ends. Let $m$, $J$ and $q$ denote the
total mass, angular momentum and charge respectively at one of the ends. Then,
the following inequality holds
\begin{equation}
  \label{eq:42}
   m^2\geq \frac{q^2+\sqrt{q^4+4J^2}}{2}.
\end{equation}
\end{theorem}
For the precise definition, fall off conditions an assumptions on the
electro-vacuum initial data we refer to \cite{Chrusciel:2009ki} and
\cite{Costa:2009hn}.  For simplicity, in \ref{sec:axially-symm-init} we discuss
in detail only the pure vacuum case.

\begin{figure}
  \centering
   \includegraphics[width=4cm]{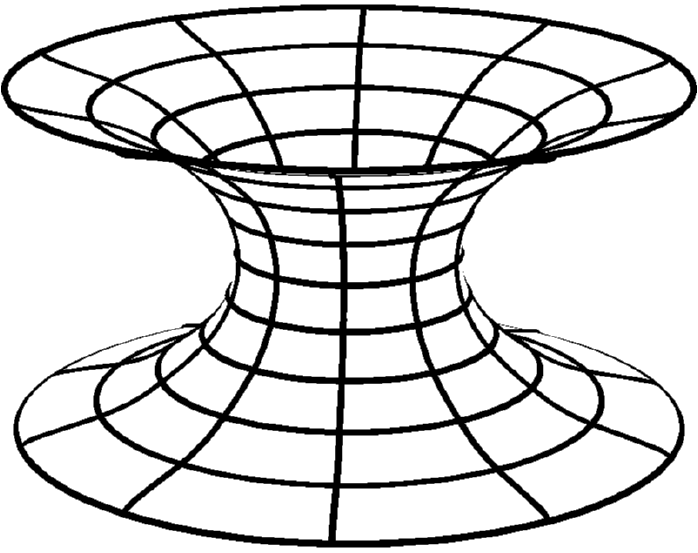}  
  \caption{Initial data with two asymptotically flat ends.}
  \label{fig:4}
\end{figure}

Recall that for asymptotically flat initial data the total mass $m$, the total
charge $q$ and the total angular momentum $J$ (without any symmetry assumption)
are well defined as integrals over two-spheres at infinity for a given
asymptotic end. That is, all the quantities involved in (\ref{eq:42}) are well
defined for generic asymptotically flat data which are not necessarily axially
symmetric. However the inequality does not hold without the symmetry
assumption. General families of counter examples have been constructed in
\cite{huang11} for pure vacuum and complete manifolds.

Under the hypothesis of this theorem (namely, electro-vacuum and axial
symmetry) both the angular momentum and the electric charge are defined as
conserved quasi-local integrals (we discuss this in detail in section
\ref{sec:angul-moment-axial}). In particular, if the topology of the manifold
is trivial (i.e. $\Rt$), then these quantities are zero and hence theorem
\ref{t:main-1} reduces to the positive mass theorem. In order to have non-zero
charge or angular momentum we need to allow non-trivial topologies, for example
manifolds with two asymptotic ends as it is the case in theorem \ref{t:main-1}
(see figure \ref{fig:4}).  An important initial data set that satisfies the
hypothesis of the theorem is provided by an slice $t=constant$ in the
non-extreme Kerr-Newman black hole in the standard Boyer-Lindquist coordinates.

This theorem has three main limitations: i) the initial data are assumed to be
maximal.  ii) there is no rigidity statement. iii) the data are assumed to have
only two asymptotic ends. Let us discuss these points in more detail. The
maximal condition plays a crucial role in the proof since it ensure a positive
definite scalar curvature.  A relevant open problem is how to remove this
condition, we will discuss it in more detail in section
\ref{sec:open-problems}.

\begin{figure}
  \centering
   \includegraphics[width=4cm]{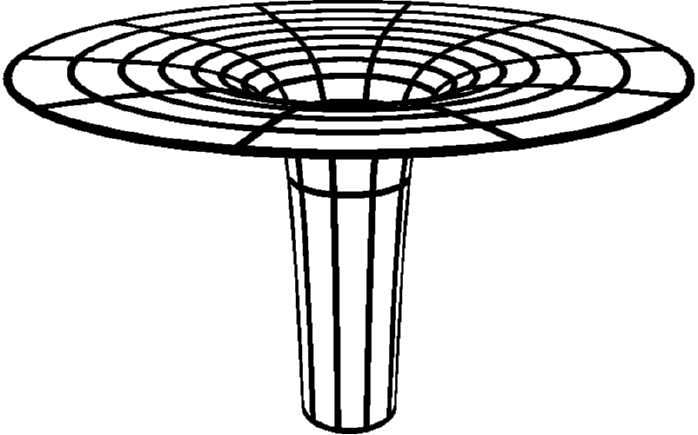}  
  \caption{The cylindrical end on extreme Kerr black hole initial data.}
  \label{fig:3}
\end{figure}

Extreme Kerr-Newman initial data, which reach the equality in (\ref{eq:42}), is
not asymptotically flat in both ends. The data have a cylindrical end and an
asymptotically flat end (see figure \ref{fig:3}). Hence these data is excluded
in the hypothesis of theorem \ref{t:main-1}. In order to include the equality
case we need to enlarge the class of data.  An example is given by the
following theorem proved in \cite{Dain06c} which includes the rigidity
statement. 

\begin{theorem}
\label{t2}
Consider a vacuum Brill initial data set such that they satisfy condition 2.5
in \cite{Dain06c}.  Then inequality
 \begin{equation}
   \label{eq:43}
   m\geq \sqrt{|J|},
 \end{equation}
holds.  Moreover, the
equality in \eqref{eq:43} holds if and only if the data are a slice
of the extreme Kerr spacetime. 
\end{theorem}
The precise definition of the Brill class of data can be seen in
\cite{Dain06c}. The main advantage of these kind of data is that they encompass
both class of asymptotics: cylindrical and asymptotic flatness.  We discuss
this in section \ref{sec:axially-symm-init}.  The condition 2.5 (see
\cite{Dain06c} for details) mentioned in this theorem implies that the initial
data have non trivial angular momentum only at one end, however multiple extra
ends with zero angular momentum are allowed.  This condition involves also
other restrictions which are technical.  In a very recent work \cite{Schoen11}
these technical conditions have been removed and also an interesting new
approach to the variational problem is presented.

The inequality (\ref{eq:42}) (which in particular implies (\ref{eq:43})) is
expected to hold for manifolds with an arbitrary number of asymptotic ends,
this generalization is probably to most important open problem regarding this
kind of inequalities (we discuss this in detail in the final section
\ref{sec:open-problems}). There exist, however, a very interesting partial
result \cite{Chrusciel:2007ak}. In order to describe it, we need to introduce
the mass functional $\mf$, this functional is defined in section
\ref{sec:mass-axial-symmetry}. It plays a major role in all the proofs, as it
is explained in section \ref{sec:main-ideas}. This functional represents a
lower bound for the mass. Moreover, the global minimum of this functional
(under appropriate boundary conditions which preserve the angular momentum) is
achieved by an harmonic map with prescribed singularities.  As we will see in
section \ref{sec:main-ideas} this is the main strategy in the proofs of all the
previous theorem which are valid for two asymptotic ends. Remarkably enough in
\cite{Chrusciel:2007ak} the existence and uniqueness of this singular harmonic
map has been proved also for manifolds with an arbitrary number of asymptotic
ends. In this article the following theorem is proved.  
\begin{theorem}
  \label{Piotr-Gilbert}
  Consider an axially symmetric, vacuum asymptotically flat and maximal initial
  data with $N$ asymptotic ends. Denote by $m_i$, $J_i$ ($i=1,\ldots N$) the
  mass and angular momentum of the end $i$. Take an arbitrary end (say $1$),
  then the mass at this end satisfies the inequality
\begin{equation}
  m_1\geq \mf(J_2,\ldots, J_N)
\end{equation}
where $ \mf(J_2,\ldots, J_N)$ denotes the numerical value of the mass
functional $\mf$ evaluated at the corresponding harmonic map. 
\end{theorem}
This theorem reduces the proof of the inequality with multiples ends to compute
the value of the mass functional on the corresponding harmonic map and verify
the inequality
\begin{equation}
  \label{eq:49}
  \mf(J_2,\ldots, J_N) \geq \sqrt{|J_1|}.
\end{equation}
We further discuss this theorem in section \ref{sec:global-inequalities-ideas}.

Strong numerical evidences that inequality (\ref{eq:49}) holds for three
asymptotic ends has been provided in \cite{Dain:2009qb}. The numerical methods
used in that article are related with the harmonic map structure of the
equations. We will describe them in section \ref{sec:harmonic-maps}.

All the previous results assume complete manifolds without inner
boundaries. The inclusion of inner boundary is important to prove the Penrose
inequality with angular momentum (\ref{eq:23}).  Boundary conditions in
relation with the mass functional $\mf$ where studied in \cite{Gibbons06} in
order to prove a version of Penrose inequality in axial symmetry.  In
\cite{Chrusciel:2011eu} inner boundaries were also included and an interesting
new lower bound for the mass is obtained which depends only on the inner
boundary. Finally, we mention that in \cite{Jaramillo:2007mi} numerical
evidences for the validity of the Penrose inequality (\ref{eq:23}) has been
presented.

\subsection{Quasi-local inequalities}
\label{sec:quasi-local-ineq}

Quasi-local inequalities between area and charge has been proved in
\cite{Gibbons:1998zr} for stable minimal surfaces on time symmetric initial
data.  The following theorem proved in \cite{Dain:2011kb} generalize this
result for generic dynamical black holes.

Consider Einstein equations with cosmological constant $\Lambda$
\begin{equation}
  \label{eq:3aa}
  G_{\mu\nu}=8\pi (T^{EM}_{\mu\nu}+T_{\mu\nu})-\Lambda g_{\mu\nu},
\end{equation}
where $T^{EM}_{\mu\nu}$ is the electromagnetic energy-momentum tensor defined
in terms of the electromagnetic field $F_{\mu\nu}$ by (\ref{eq:27}). The
electric charge of an arbitrary closed, oriented, two-surface $\Su$ embedded in
the spacetime is defined by (\ref{eq:26}).

\begin{theorem}
\label{t:Aq}
Given a closed marginally trapped surface ${\cal S}$ satisfying spacetime
stably outermost condition, in a spacetime which satisfies Einstein equations
(\ref{eq:3aa}) with non-negative cosmological constant $\Lambda$ and such that
the non-electromagnetic matter fields $T_{\mu\nu}$ fulfill the dominant
energy condition, the following inequality holds:
\begin{equation}
\label{e:inequalityq}
A \geq 4\pi q^2 ,
\end{equation}
where $A$ and $q$ are the area and the charge of ${\cal S}$.
\end{theorem}
For the definition of marginally trapped surfaces (which is standard) and the
stably condition see \cite{Andersson:2005gq} \cite{andersson08}
\cite{Jaramillo:2011pg} \cite{Dain:2011kb} (see also \cite{Hayward93}
\cite{Racz:2008tf}).  This theorem is a completely quasi-local result that
applies to general dynamical black holes without any symmetry assumption. It is
also important to emphasize that the matter is not assumed to be uncharged,
namely it is allowed that $\nabla_\mu F^{\mu\nu}\neq 0$ (which is equivalent to
$\nabla^\mu T^{EM}_{\mu\nu}\neq 0$ ).  The only condition imposed in the
non-electromagnetic matter field stress-energy tensor $T_{\mu\nu}$ is that it
satisfies the dominant energy condition.   In this theorem it is also possible to
include the magnetic charge  and Yang-Mills charges  (see \cite{Dain:2011kb}
and  \cite{Jaram12}).

In \cite{Simon:2011zf} an interesting generalization of theorem \ref{t:Aq} is
presented in which the stability requirement is removed at the expense of
introducing the principal eigenvalue of the stability operator and also the
cosmological constant (with arbitrary sign) is added. 

At the end of section \ref{sec:physical-picture} we have observed that a
variant of inequality (\ref{e:inequalityq}) is expected to hold for ordinary
macroscopic charged object (which are not necessarily black holes) at least if
they are `round enough'.  In fact there exists an interesting and highly
non-trivial counter example to (\ref{e:inequalityq}) for macroscopic
objects. This counter example was constructed by W. Bonnor in \cite{bonnor98}
(see also the discussion in \cite{Dain:2011kb}) and it can be summarized as
follows: for any given positive number $k$, there exist static, isolated,
non-singular bodies, satisfying the energy conditions, whose surface area $A$
satisfies $A\leq k q^2$ .  The body is a highly prolated spheroid of
electrically counterpoised dust.  From the physical point of view we are saying
that for an ordinary charged object (in contrast to a black hole) we need to
control another parameter (the ‘roundness’) in order to obtain an inequality
between area and charge.  Remarkably enough it is possible to encode this
intuition in the geometrical concept of isoperimetric surface: we say that a
surface $\Su$ is isoperimetric (or `round') if among all surfaces that enclose
the same volume as $\Su$ does, $\Su$ has the least area. Then, based on the
results proved in \cite{christodoulou88} the following theorem was obtained in
\cite{Dain:2011kb} for isoperimetric surfaces.
\begin{theorem}
\label{t:main3} 
Consider an electro-vacuum, maximal initial data, with a non-negative
cosmological constant. Assume that $\Su$ is a stable isoperimetric sphere. Then
\begin{equation}
  \label{eq:isoparametric} 
A \geq \frac{4\pi}{3} q^{2}, 
\end{equation}
where  $q$  is the electric  charge of
$\Su$.   
\end{theorem}
Note that inequality (\ref{eq:isoparametric}) has a different coefficient as
(\ref{e:inequalityq}). We recall that the notion of stable isoperimetric surface is very
similar to the case of a stable minimal surface: the differential operator is
identical, the only difference is that the allowed test functions should
integrate to zero on the surface, this is precisely the condition that the
deformations preserve the volume (see \cite{Barbosa88}).  

It is also possible to prove interesting variants of theorem \ref{t:Aq} which
are valid for generic surfaces (i.e. not necessarily trapped or minimal) but in
order to obtain these results global assumptions on the initial data should be
made (that is, in contrast to theorem \ref{t:Aq}, these are not a purely
quasi-local results): the two surfaces are embedded on initial conditions that
are complete, maximal and asymptotically flat. The non electromagnetic matter
fields are assumed to be non charged and they should satisfy the dominant
energy condition on the whole initial data (see Theorem 2.2 in
\cite{Dain:2011kb}).

As in the case of the global inequality, quasi-local inequalities with angular
momentum involve different techniques in comparison with the pure charged
case. Their study star very recently. 
 
The quasi-local inequality with angular momentum and charge (\ref{eq:5}) was
first conjectured to hold in stationary spacetimes in \cite{Ansorg:2007fh}. In
that article the extreme limit of this inequality was analyzed and also
numerical evidences for the validity in the stationary case was presented
(using the numerical method and code developed in \cite{Ansorg05}). In a series
of articles \cite{hennig08} \cite{Hennig:2008zy} the inequality (\ref{eq:5})
was proved for stationary black holes.  See also the review article
\cite{Ansorg:2010ru}.

It is important to emphasize that the stationary non-vacuum case is highly
non-trivial. The physical situation is, for example, a black hole surrounded by
a ring of matter (which do not touch the black hole). To illustrate the
complexity of this case we mention that the Komar mass (which is only defined
in the stationary case) can be negative for these black holes (for the
Kerr-Newman black hole is always positive), see \cite{Ansorg:2006qt}
\cite{Ansorg:2007vt}.  It is interesting to mention that for this
class of stationary spacetimes there exists a remarkable relation of the form
$(8\pi J)^2+ (4\pi q^2)^2=A^+A^-$, where $A^+$ and $A^-$ denote the areas of
event and Cauchy horizon. This result have been proved in the following series
of articles \cite{Ansorg:2009yi} \cite{Hennig:2009aa} \cite{Ansorg:2008bv}. 

In the dynamical regime, the inequality was conjectured to hold in
\cite{dain10d} based on the heuristic argument mentioned in section
\ref{sec:physical-picture}. In that article also the main relevant techniques
for its proof were introduced, namely the mass functional on the surface and
its connections with the area (we discuss this in section
\ref{sec:two-surfaces-mass}). A global proof (but with technical restrictions)
was obtained in \cite{Acena:2010ws} \cite{Clement:2011kz}. The first general
and pure quasi-local result was proven in \cite{Dain:2011pi}, where the
relevant role of the stability condition for minimal surfaces was pointed out:
\begin{theorem}
\label{t:main-2}
Consider an axisymmetric, vacuum and maximal initial data, with a non-negative
cosmological constant. Assume that the initial data contain
an orientable closed stable minimal axially symmetric surface
$\Su$. Then
\begin{equation}
\label{desigualdad}
 A \geq 8\pi |J|,
\end{equation}
where $A$ is the area and $J$ the angular momentum of $\Su$. Moreover, if
the equality in \eqref{desigualdad} holds then $\Lambda=0$ and the local
geometry of the surface $\Su$ is an extreme Kerr throat sphere.
\end{theorem}
The extreme throat sphere geometry, with angular momentum $J$, was defined in
\cite{dain10d} (see also \cite{Acena:2010ws} and \cite{Dain:2011pi}). This
surface capture the local geometry near the horizon of an extreme Kerr black
hole and it is defined as follows. The sphere is embedded in an initial
data with intrinsic metric given by
\begin{equation}
\label{eq:gamma0}
\gamma_0=4J^2e^{-\sigma_0}d\theta^2+ e^{\sigma_0}\sin^2\theta d\phi^2,
\end{equation}
where $\sigma_0$ is given by (\ref{datos}).
Moreover, the sphere must be totally geodesic, the twist
potential evaluated at the surface must be given by $\omega_0$ defined by (\ref{datos}) 
and  the  components of the second fundamental 
\begin{equation}
K_{ij}\xi^i= K_{ij}n^j n^i= K_{ij}\eta^j \eta^i = 0,
\end{equation}
must vanish at the surface. Here $K_{ij}$ denotes the second fundamental form
of the initial data, $n^i$ the unit normal vector to the surface and $\eta^i$
the axial Killing field. Note that the functions $\sigma_0$ and $\omega_0$,
which characterize the intrinsic and extrinsic geometry of the surface
respectively, depend only on the angular momentum parameter $J$.  The geometry
of axially symmetric initial data set are described in detail in section
\ref{sec:axially-symm-init}. In particular, the twist potential is determined
by the second fundamental form $K_{ij}$, using equation (\ref{eq:44}). The
adapted coordinates system used in (\ref{eq:gamma0}) is defined is section
\ref{s:potentials}.

This theorem has two main restriction: the first one is the maximal
condition. The second is vacuum. Remarkably enough, it is possible not only to
avoid both restrictions but also to provide a pure spacetime proof (that is, no
mention of a three-dimensional hypersurface) of this inequality, in which
axisymmetry is only imposed on $\Su$. This generalization is proved in
\cite{Jaramillo:2011pg}:
\begin{theorem}
  Given an axisymmetric closed marginally trapped surface $\Su$ satisfying
  the (axisymmetry-compatible) spacetime stably outermost condition, in a
  spacetime with non-negative cosmological constant and fulfilling the dominant
  energy condition, it holds the inequality
\begin{equation}
\label{e:inequality}
A \geq 8\pi |J|, 
\end{equation}
where $A$ and $J$ are the area and (Komar) angular momentum of $\Su$. If
equality holds, then $\Su$ is a section of a non-expanding  horizon with
the geometry of extreme Kerr throat sphere.
\end{theorem}
The concept of non-expanding horizon is explained in \cite{Jaramillo:2011pg},
it essentially means that the shear vanished at the surface.  

It is important to note that the angular momentum that appears in
(\ref{e:inequality}) is the gravitational one (i.e. the Komar integral). The
matter fields have also angular momentum and it can be transferred to the black
hole, however the inequality (\ref{e:inequality}) remains true even in that
case. In fact this inequality is non-trivial for the Kerr-Newman black hole, we
discuss this in detail in section \ref{sec:angul-moment-axial}.

In \cite{reiris11} it has been pointed out that there exists a connexion
between the global inequalities described in section
\ref{sec:global-inequality} and the quasi-local inequalities. This is obtained
by linking the relevant mass functional $\mf$ and $\mfs$ (we discuss these mass
functional in section \ref{sec:main-ideas}).

Inequality (\ref{e:inequality}) has played an important role in the proofs of
the non-existence of stationary two black holes configurations
(see \cite{Neugebauer:2011qb} \cite{Chrusciel:2011iv}).

Finally, we mention that there exists very interesting generalization of
(\ref{e:inequality}) to black holes in higher dimensions
\cite{Hollands:2011sy}.

\subsection{Main ideas}
\label{sec:main-ideas}
In this section we present the main ideas behind the proofs of the global
inequality (\ref{eq:42}) and the quasi-local inequality
(\ref{e:inequality}). This section should be consider as a guide in which the
technicalities are avoided. In section \ref{sec:angul-moment-axial} and
\ref{sec:mass-axial-symmetry} we discuss in details the main relevant
properties of the angular momentum and mass in axial symmetry which constitute
the essential part of the proofs.

\subsubsection{Global inequalities}
\label{sec:global-inequalities-ideas}

The starting point in the proof of the inequality (\ref{eq:42}) (we consider
the case $q=0$ for simplicity) is the formula for the total mass $m$ given by
Eq. (\ref{eqq:103}). This formula represents the total mass as a positive
definite integral over a maximal (here is where the maximal condition plays a
crucial role) initial surface. This integral representation is a generalization
of the Brill mass formula discovered in \cite{Brill59} (in section
\ref{sec:mass-axial-symmetry} we further discuss this formula and provide the
relevant references). The formula holds in a particular coordinate system which
is called isothermal (see lemma \ref{isothermal}).  
 
The integrand in Eq. (\ref{eqq:103}) has two kind of terms: dynamical and
stationary. The dynamical terms vanished for an stationary solution like
Kerr. The stationary part lead to the relevant mass functional $\mf$ defined by
(\ref{eqq:5c}). This functional provides an obvious lower bound to the total
mass, namely
\begin{equation}
  \label{eq:2}
  m\geq \mf(\sigma, \omega).
\end{equation}
The mass functional $\mf$ depends on two functions $\sigma$ and $\omega$. The
function $\sigma$ is essentially the norm of the axial Killing vector (see
equation (\ref{eq:83})). The function $\omega$ is the twist potential of the
Killing vector (see section \ref{sec:angul-moment-axial}).  These two functions
can be freely prescribed on the initial data. This is an important and far from
obvious property since the constraint equations (\ref{const1})--(\ref{const2})
should be satisfied.  This fact allows to formulate a variational principle for
the functional $\mf$. The other important ingredient for this variational
principle is the behavior of the angular momentum.  The angular momentum is
prescribed by the value of $\omega$ at the axis (see section
\ref{sec:angul-moment-axial}). Then, if the variations of $\omega$ vanishes at
the axis the angular momentum will be preserved. With this two ingredients, it
is hence possible to reduce the proof of the inequality  (\ref{eq:42}) to a
pure variational problem for the functional $\mf(\sigma, \omega)$. 

The second step of the proof is to solve this variational problem. This is the
most difficult part and also the most interesting since it reveals the
geometric properties of the mass functional $\mf$. Let us discuss the general
strategy of the proof. 

Let $(\sigma_0, \omega_0)$ be the corresponding functions obtained from the
extreme Kerr initial data with angular momentum $J$, then we have that
\begin{equation}
  \label{eq:3}
  \mf(\sigma_0, \omega_0)=\sqrt{|J|}.
\end{equation}
The heuristic discussed in section \ref{sec:physical-picture} and
Eq. (\ref{eq:10}) suggest that the following inequality holds
\begin{equation}
  \label{eq:10}
   \mf(\sigma, \omega)\geq \sqrt{|J|},
\end{equation}
for all $(\sigma, \omega)$ such that $\omega$ has the same value at the axis as
the function $\omega_0$. Moreover, the equality in (\ref{eq:10}) is reached if
and only if $\sigma=\sigma_0$ and $\omega=\omega_0$.  This is precisely the
variational problem. Note the variational problem is formulated purely in terms
of the functional $\mf(\sigma, \omega)$, without any reference to the
constraint equations (\ref{const1})--(\ref{const2}).

The first evidence that this variational problem will have the expected
solution is that the Euler-Lagrange equations of the functional $\mf$ are
equivalent to the stationary axially symmetric Einstein equations, in
particular extreme Kerr satisfies these equation \cite{Dain05c}. The second
evidence (which is harder to prove) is that the second variation of $\mf$ is
positive definite evaluated at extreme Kerr \cite{Dain05d}. To prove this
positivity property it is crucial to make contact with the harmonic map theory
(in this case, trough the Carter identity). With these ingredients it is
possibly to show that extreme Kerr is a local minimum of the mass and hence the
inequality (\ref{eq:10}) is proved in an appropriate defined neighborhood of
extreme Kerr. This local proof was done in \cite{Dain05d}.

To have a global proof of (\ref{eq:10}) (i.e. without any smallness assumption)
more subtle properties of the mass functional are required. A crucial step is
to realize that the mass functional $\mf$ is essentially the renormalized
energy of an harmonic map into the hyperbolic plane \cite{Dain06c} (we discuss
this in section \ref{sec:harmonic-maps}). This kind of harmonic maps have been
extensive studied in the literature. The problem here is that the map is
singular at the axis and hence the standard techniques do not apply directly.
To use the harmonic map theory we need to handle these singularities and this
is the main technical difficulty.  In \cite{Dain06c} the proof of this
variational result was done using estimates, inspired in the work of
G. Weinstein \cite{Weinstein90} \cite{Weinstein92} \cite{Weinstein94}
\cite{Weinstein95} \cite{Weinstein96} \cite{Weinstein96b} (see also
\cite{Li92})), which rely on particular properties of this functional
(inversion symmetry). In subsequent works \cite{Chrusciel:2007dd},
\cite{Chrusciel:2007ak}, \cite{Chrusciel:2009ki}, \cite{Costa:2009hn} the proof
was simplified and improved using general results on harmonic maps (more
precisely the existence result \cite{Hildebrandt77}). In these proofs the
connection with the harmonic maps theory is more transparent and the problem of
the singularities is clearly isolated.

Finally let us mention the following important point regarding initial data
with multiples ends. The multiples ends appears in the variational problem
(\ref{eq:10}) as singular points of the functions $(\sigma, \omega)$.
Remarkably, even in that case it is possible to solve completely the
variational problem. This is precisely the content of theorem
\ref{Piotr-Gilbert} (proved in \cite{Chrusciel:2007ak}). The only missing piece
in order to prove the inequality (\ref{eq:42}) in that case is the
following. Theorem \ref{Piotr-Gilbert} ensure the existence of a global minimum
but the value of $\mf$ at the global minimum is unknown. In the case of two
asymptotic ends we known that extreme Kerr is the global minimum and hence we
can explicitly compute the value (\ref{eq:3}).

\subsubsection{Quasi-local inequalities}
\label{sec:quasi-local-ineq-ideas}
The global inequality (\ref{eq:42}) applies to a complete three dimensional
manifold. In contrast the quasi-local inequality (\ref{e:inequality}) apply to
a closed two-surface. In principle it is a priori not clear at all that there
is a relation between these two kind of inequalities.  The two physical
heuristic argument presented in section \ref{sec:physical-picture} in support
for them are very different.  In particular, it is far from obvious that the
mass functional $\mf$ can play a role for the quasi-local
inequalities. Remarkably enough, it turns out that a suitable adapted mass
functional $\mfs$ over a two-surface play a very similar role as $\mf$.  The
motivation for the definition of $\mfs$ given by (\ref{eqq:44}) is discussed in
section \ref{sec:two-surfaces-mass}. The new mass functional $\mfs$ and its
connection with the area represent the kernel of the proof of
(\ref{e:inequality}). Let us discuss this.

On the extreme Kerr initial data there exist an important canonical
two-surface, namely the intersection of the Cauchy surface with the horizon. On
a surface given by $t=constant$ in the Boyer-Lindquist coordinates this
two-surface is located at infinity on the cylindrical end (see figure
\ref{fig:3}). Let call $(\bar \sigma_0, \bar \omega_0)$ the value of the
functions $(\sigma_0, \omega_0)$ (defined in the previous section
\ref{sec:global-inequalities-ideas}) on this two-surface.  The functions $(\bar
\sigma_0, \bar \omega_0)$ will play for $\mfs$ a similar role as the functions
$(\sigma_0, \omega_0)$ for $\mf$. Namely, first they satisfy the Euler-Lagrange
equations of $\mfs$. Second, the second variation of $\mfs$ is positive
definite evaluated at $(\bar \sigma_0, \bar \omega_0)$ (see \cite{dain10d}).
There is also a connection with the harmonic maps energy (we discuss this in
section \ref{sec:two-surfaces-mass}). Using similar kind of arguments as
described in section \ref{sec:global-inequalities-ideas} it is possible to
prove the following inequality 
\begin{equation}
  \label{eq:12}
  2|J|\leq e^{(\mfs(\sigma,\omega)-8)/8}
\end{equation}
for all functions $(\sigma, \omega)$ such that $\omega$ has the same values at
the poles of the two-surface as $\omega_0$. With equality if and only if we
have $\sigma=\bar \sigma_0$ and $\omega=\omega_0$. A local version of this
inequality was first proved in \cite{dain10d}. The global version (\ref{eq:12})
was proved in \cite{Acena:2010ws}. The rigidity statement was proved in
\cite{Dain:2011pi}. We emphasize that the inequality (\ref{eq:12}) (in complete
analogy to the inequality (\ref{eq:10})) is a property of the functional
$\mfs$, the geometry of the two-surface $\Su$ does not intervene at all.
 
The inequality (\ref{eq:12}) is interesting but, in the light of the discussion
presented in \ref{sec:global-inequalities-ideas}, it is somehow expected.  What
is crucial and completely unexpected is the relation between $\mfs$ and the
area of the surface $\Su$.  This relation was founded locally in \cite{dain10d}
and globally in \cite{Acena:2010ws}. By local in this context we mean that the
relation holds for surfaces in an appropriated defined neighborhood of an
extreme Kerr throat geometry. On the other hand, global means that the relation
holds for general surfaces.  In \cite{Dain:2011pi} the important connection
with the stability condition was proved for minimal surfaces and in
\cite{Jaramillo:2011pg} for marginally trapped surfaces. The final inequality
essentially reads as follows. For a two-surface $\Su$ such that: (i) it is
minimal \cite{Dain:2011pi} (or marginally trapped \cite{Jaramillo:2011pg}) and
(ii) it is stable, then the following inequality holds
\begin{equation}
  \label{eq:18}
  A\geq 4\pi e^{(\mfs-8)/8}.
\end{equation}
The minimal and marginally trapped conditions are requirements on the extrinsic
curvature of $\Su$ (namely, on the trace of the second fundamental form).  The
stability condition is a requirement on derivatives of the second fundamental
form. In the spacetime version presented in \cite{Jaramillo:2011pg}) the
inequality \eqref{eq:18} is a consequence of a flux inequality (see Lemma 1 in
that reference) where the geometric and physical meaning of each term is
apparent.

\section{Angular momentum in axial symmetry}
\label{sec:angul-moment-axial}
Axial symmetry plays a major role in the inequalities that include angular
momentum presented in the previous sections for two main reasons. For the
global inequality (\ref{eq:42}) is the conservation of angular momentum implied
by axial symmetry which is relevant. For the quasi-local inequality
(\ref{desigualdad}), is the very definition of quasi-local angular momentum
(only possible in axial symmetry) which is important. These two properties are
closed related for vacuum spacetimes, since the Komar integral provides both
the conservation law and the definition of quasi-local angular momentum. For
non-vacuum spacetimes in axial symmetry the Komar integral still provides a
meaningful expression for the gravitational angular momentum but it is no
longer conserved. Nevertheless, as we already mentioned in section
\ref{sec:physical-picture}, in the electro-vacuum case the sum of the
gravitational and electromagnetic angular is conserved, and hence in that case
is possible to prove the global inequality (\ref{eq:42}). In the general
non-vacuum case in axial symmetry (that is, with general matter sources which
are not electromagnetic) no simple and universal relation between mass and
angular momentum is expected. For example in \cite{Giacomazzo:2011cv} neutron
starts models in axial symmetry have been numerically constructed such that
they violate inequality (\ref{eq:42}).  Nevertheless, remarkably, the
quasi-local inequality (\ref{desigualdad}) for black holes is still valid in
the general non-vacuum axially symmetric case.

In this section we summarize relevant results concerning angular momentum in
axial symmetry. Although these results are not new they are not easy to find in
the literature, notably, the conservation of angular momentum in the
electro-vacuum case and the relation between Komar integrals and potentials in
the presence of matter fields. The conservation of angular momentum in the
electro-vacuum, together with the relevant Komar integral for the
electromagnetic angular momentum, were discovered in \cite{Simon:1984qb}, in a
much general setting. We present a simpler derivation of this result in the
language of differential forms.

Let us begin with some general remarks about conserved quantities in General
Relativity (see also the review article \cite{jaramillo11c} for a related
discussion and \cite{Szabados04} for the general problem of how to define
quasi-local angular momentum without symmetries).

Let $M$ be a four dimensional manifold with metric $g_{\mu\nu}$ (with signature
$(-+++)$) and Levi-Civita connection $\nabla_\mu$. On this curved background,
let us consider an arbitrary energy-momentum tensor $T_{\mu\nu}$ which
satisfies the conservation equation
\begin{equation}
  \label{eqcm:4}
  \nabla_\mu T^{\mu\nu}=0.
\end{equation}
It is well known that if the spacetime admit a Killing vector field $\eta^\mu$
\begin{equation}
  \label{eqcm:12}
  \nabla_{(\mu} \eta_{\nu)}=0,
\end{equation}
then the vector
\begin{equation}
  \label{eqcm:5}
 K_{\mu}= T_{\mu \nu}\eta^\nu,
\end{equation}
is divergence free 
\begin{equation}
  \label{eq:64}
  \nabla_\mu  K^{\mu}=0.
\end{equation}
This equation provides an integral conservation law via Gauss theorem.  For
some of the computations in this section it is convenient to use differential
forms instead of tensors. We will denote them with boldface.  Let
$\boldsymbol{K}$ be the 1-form defined by (\ref{eqcm:5}). Equation
(\ref{eq:64}) is equivalent to
 \begin{equation}
   \label{eq:63}
  d{}^*\boldsymbol{K}=0, 
 \end{equation}
 where $d$ is the exterior derivative and the dual of a $p$ form is defined
 with respect to the volume element $\epsilon_{\mu\nu\lambda\gamma}$ of the
 metric $g_{\mu\nu}$ by the standard formula
\begin{equation}
  \label{eq:61}
  {}^*\alpha_{\mu_1\cdots \mu_{4-p}}=\frac{1}{p!}\alpha^{\nu_1\cdots \nu_p}
    \epsilon_{\nu_1\cdots \nu_p \mu_1\cdots  \mu_{4-p}}.
\end{equation}
Let $\Omega$ denotes a four-dimensional, orientable, region in $M$ and let
$\partial \Omega$ be its three-dimensional boundary.  Then using (\ref{eq:63})
and the Stokes theorem we obtain
\begin{equation}
  \label{eqcm:22}
 0= \int_\Omega d{}^*\boldsymbol{K}=  \int_{\partial \Omega} {}^*\boldsymbol{K}.
\end{equation}
Note that in this equation the region $\Omega$ is arbitrary and the boundary
$\partial \Omega$ can have many disconnected components.

Consider a spacelike three-surface $S$.  The conserved quantity corresponding
to the Killing vector $\eta^\mu$ is defined with respect to $S$ by
\begin{equation}
  \label{eqcm:8}
  K(S)=  \int_{S}  {}^*\boldsymbol{K}. 
\end{equation}
The interpretation of equation (\ref{eqcm:22}) in relation to the quantity
$K(S)$ is the following.  Let $\Omega$ to be a timelike cylinder, such that its
boundary $\partial \Omega$ is formed by the bottom and the top spacelike
surfaces $S_1$ and $S_2$ and the timelike piece $\mathcal{C}$. Then we have
(taking the corresponding orientation)
\begin{equation}
  \label{eqcm:7}
  0= \int_{\partial \Omega} {}^*\boldsymbol{K}=  K(S_1) - K(S_2) +
  \int_{\mathcal{C}} {}^*\boldsymbol{K}. 
\end{equation}
The integral over the timelike surface $\mathcal{C}$ is the flux of $K(S)$.
Equation (\ref{eqcm:7}) is interpreted as the conservation law of the quantity
$K(S)$.  The region $\Omega$ can also be chosen to have a null boundary
$\mathcal{C}$, equation (\ref{eqcm:7}) remains identical and the interpretation
is similar.

If the Killing vector $\eta^\mu$ is also a symmetry of the tensor $T_{\mu\nu}$
(we have not assumed that so far), namely
\begin{equation}
  \label{eqcm:10}
  \pounds_\eta T_{\mu\nu}= 0,
\end{equation}
where $\pounds$ denote Lie derivatives, then the following vector is also
divergence free 
\begin{equation}
  \label{eqcm:11}
 \hat K_\mu= 8\pi (T_{\mu \nu}\eta^\nu -\frac{1}{2} T \eta_{\mu}),
\end{equation}
where $T$ is the trace of $T_{\mu \nu}$. In fact there is a whole family of
divergence free tensors since $T\eta_\mu$ is divergence free. Hence the
previous discussion applies to this vector as well. 

Note that the conserved quantities $K(S)$ are naturally defined as integrals
over spacelike three-surfaces. In flat spacetime it is possible to convert
these integrals into a boundary integral over two-surfaces. This gives the
quasi-local representation of conserved quantities (see the discussion in the
introduction of \cite{Szabados04}). In a curved background this is in general
not possible.  However, as we will see, this possible for the particular case
of the electromagnetic field.

Before analyzing the angular momentum it is important to study the electric
charge. The electric charge is of course relevant in our discussion since it
appears in the inequalities discussed in the previous  sections. But
more important, even if we want to analyze these inequalities in pure vacuum,
the electric charge represent the simpler `conserved charge' on a curved
spacetime. Its definition and properties serve as model for all the other
conserved quantities, like the angular momentum.

The Maxwell equations on $(M, g_{\mu\nu})$ are given by
\begin{align}
  \label{eq:25}
  \nabla^\mu F_{\mu\nu} & =-4\pi j_\nu, \\ 
\nabla_{[\mu} F_{\nu \alpha]} & =0.
\end{align}
In terms of forms, they are written as
\begin{align}
  \label{eq:51}
  d {}^* \mathbf{F} &=4\pi {}^* \mathbf{j},\\
d \mathbf{F} &=0.
\end{align}
The energy-momentum tensor of the electromagnetic field is given by
\begin{equation}
  \label{eq:27}
  T_{\mu\nu}=\frac{1}{4\pi}\left(F_{\mu\lambda}
    F_\nu{}^{\lambda}-\frac{1}{4}g_{\mu\nu} F_{\lambda\gamma} F^{\lambda\gamma}  \right).
\end{equation}

Let $\Su$ be  a closed orientable two-surface  embedded in $M$ (in the
following, all two-surfaces will be assumed to be closed and orientable). 
The electric charge $q$ of $\Su$ is defined by 
\begin{equation}
  \label{eq:26}
  q(\Su)=\frac{1}{4\pi}\int_{\Su} {}^* \mathbf{F}.
\end{equation}
Let $S$ be a three-surface with boundary $\Su$, then using Stokes theorem
and Maxwell equation (\ref{eq:51}) we obtain
\begin{equation}
  \label{eq:28}
  q(\Su)=    \int_S   {}^* \mathbf{j}.
\end{equation}
This equation is interpreted as follows. 
From equation (\ref{eq:51}) we deduce the conservation law for the current
$\mathbf{j}$ analog to (\ref{eq:63}), namely
\begin{equation}
  \label{eq:19}
 d {}^* \mathbf{j}=0. 
\end{equation}
And hence taking the same region $\Omega$ and using Stokes theorem we obtain
the analog expression as (\ref{eqcm:7}) for the current $\mathbf{j}$
\begin{equation}
  \label{eq:29}
  0=  \int_{S_1}   {}^* \mathbf{j}-  \int_{S_2}   {}^* \mathbf{j}+
  \int_{\mathcal{C}}   {}^* \mathbf{j}. 
\end{equation}
Using (\ref{eq:28}) we finally obtain
\begin{equation}
  \label{eq:30}
  q(\Su_1)- q(\Su_2)= \int_{\mathcal{C}}   {}^* \mathbf{j}. 
\end{equation}
This is the conservation law for the electric charge. Note that in the left
hand side of (\ref{eq:30}) we have integrals over two-surfaces, in contrast
with (\ref{eqcm:7}) where integrals over three-surfaces appear. This is because
we have an extra equation (i.e. Maxwell equation (\ref{eq:51})) that allow us
to write the integral (\ref{eq:28}) in the form (\ref{eq:26}).

When $\mathbf{j}=0$ the charge has the same value, namely
\begin{equation}
  \label{eq:62}
  q(\Su_1)= q(\Su_2), 
\end{equation}
and we say that the charge is strictly conserved.

We turn now to angular momentum for axially symmetric spacetimes. We begin with
the definition of axial symmetry. 
\begin{definition}
\label{d:ax}
  The spacetime $(M,g_{\mu\nu})$ is said to be axially symmetric if its group of
  isometries has a subgroup isomorphic to $SO(2)$.  
\end{definition}
We will denote by $\eta^\mu$ the Killing field generator of the axial
symmetry. The orbits of $\eta^\mu$ are either points or circles. The set of
point orbits $\Gamma$ is called the axis of symmetry. Assuming that $\Gamma$ is
a surface, it can be proved that $\eta^\mu$ is spacelike in a neighborhood of
$\Gamma$ (see \cite{Mars:1992cm}). We will further assume that the Killing
vector is always spacelike outside $\Gamma$. Note that if this condition is not
satisfied then the spacetime will have closed causal curved, in particular it
can not be globally hyperbolic.

The form $\eta_\mu$ will be denoted by $\boldsymbol{\eta}$, and the square of
its norm by $\eta$, namely
\begin{equation}
  \label{eq:7}
  \eta=\eta^\mu\eta_\mu=|\boldsymbol{\eta}|^2.
\end{equation}
We have used the notation $\eta^\mu$ to denote the Killing vector field and
$\eta$ to denote the square of its norm to be consistent with the
literature. However, in this section, to avoid confusions between $\eta^\mu$
and its square norm $\eta$, we will denote the vector field $\eta^\mu$ by
$\kif$ in equations involving differential forms in the index free notation.

Consider now  Einstein equations on an axially symmetric spacetime
\begin{equation}
  \label{eqcm:2}
  G_{\mu\nu}=8\pi T_{\mu\nu}.
\end{equation}
Note that the Killing equation (\ref{eqcm:12}) implies that $T_{\mu\nu}$
satisfies (\ref{eqcm:10}).

The Komar integral (it is also appropriate to call it the Komar charge) of the
Killing field is defined over a two-dimensional surface $\Su$ as follows
\begin{equation}
  \label{eqcm:1}
  J(\Su)=\frac{1}{16\pi}\int_\Su
  \epsilon_{\mu\nu\lambda\gamma}\nabla^\lambda \eta^\gamma= \frac{1}{16\pi}
  \int_\Su {}^* d \boldsymbol{\eta}.
\end{equation}
Hence, as we discussed above, via Stokes theorem we obtain 
\begin{equation}
  \label{eqcm:52}
  J(\Su)= \frac{1}{16\pi}
  \int_S d {}^* d \boldsymbol{\eta},
\end{equation}
where $S$ is a three-dimensional surface with boundary $\Su$. 
It is a classical result \cite{Komar59} (see also \cite{Wald84}) that the
integrand   in (\ref{eqcm:52}) can be computed in terms of the Ricci tensor 
\begin{equation}
  \label{eqcm:9}
  \nabla_{[\mu} \left(\epsilon_{\nu\alpha] \beta\gamma}
  \nabla^{\beta}\eta^{\gamma}\right)=\frac{2}{3}R^{\mu}{}_\nu\eta^\nu
\epsilon_{\mu\alpha  \beta\gamma}. 
\end{equation}
In terms of forms this equation is written as 
\begin{equation}
  \label{eqcm:13}
  d  {}^*d \boldsymbol{\eta}=  2 {}^* \boldsymbol{K},
\end{equation}
where we have defined the 1-form $\boldsymbol{K}$ by
\begin{equation}
  \label{eqcm:14}
\boldsymbol{K}\equiv  K_\mu =R_{\mu\nu}\eta^\nu.
\end{equation}
Using Einstein equations (\ref{eqcm:2}) the form $\boldsymbol{K}$ can be
written in terms of the energy momentum tensor
\begin{equation}
  \label{eqcm:15}
   K_\mu =8\pi (T_{\mu\nu}\eta^\nu-\frac{1}{2}T \eta_\mu) .
\end{equation}
Note that this expression is identical to (\ref{eqcm:11}).  Then, repeating the
same argument, we obtain the conservation law for angular momentum in axial
symmetry which is the exact analog to the charge conservation (\ref{eq:30})
\begin{equation}
  \label{eqcm:41}
J(\Su_1)- J(\Su_2)=\frac{1}{8\pi} \int_{\mathcal{C}}   {}^* \mathbf{K}. 
\end{equation}
The right hand side of this equation represent the change in the angular
momentum of the gravitational field which is produced by the left hand side,
namely the angular momentum of the matter fields. Note that the angular
momentum of the matter fields are written as integrals over three-dimensional
surfaces. In particular in vacuum we have the strict conservation of angular momentum
\begin{equation}
  \label{eq:68}
 J(\Su_1)=J(\Su_2). 
\end{equation}

We consider now the case where the energy-momentum tensor in Einstein equations
(\ref{eqcm:2}) is given purely by the electromagnetic field (\ref{eq:27}). For
simplicity we consider the case with no currents $j=0$, which is the relevant
one since only in that case we get conserved quantities.  In that case we have
that (\ref{eq:27}) satisfies the equation (\ref{eqcm:4}) and the source-free
Maxwell equations are given by
\begin{align}
  \label{eqcm:51}
  d {}^* \boldsymbol{F} &= 0,\\
d \boldsymbol{F} &=0. \label{eqcm:51b}
\end{align}
We also assume  that the Maxwell
fields are  axially symmetric, namely
\begin{equation}
  \label{eqcm:29}
  \pounds_\eta \boldsymbol{F}=0. 
\end{equation}
Consider the 1-forms defined by 
\begin{equation}
  \label{eqcm:30}
  \boldsymbol{\alpha}=\boldsymbol{F}(\kif) , \quad  \boldsymbol{\beta}={}^*
  \boldsymbol{F}(\kif),
\end{equation}
where we have used the standard notation
$\boldsymbol{F}(\kif)=F_{\mu\nu}\eta^\mu$ to denote contractions of forms with
vector fields. Using the general expression for the action of the Lie
derivative on forms
\begin{equation}
  \label{eqcm:42}
  \pounds_{\kif} \boldsymbol{\omega} =d [ \boldsymbol{\omega}(\kif)]+ (d
  \boldsymbol{\omega})(\kif),
\end{equation}
Maxwell equations  (\ref{eqcm:51})--(\ref{eqcm:51b}) and the condition 
(\ref{eqcm:29}) we obtain 
\begin{align}
  \label{eqcm:31}
 d \boldsymbol{\alpha}  = 0, \quad  d \boldsymbol{\beta}=0.
\end{align}
It follows that there exist locally  functions $\chi$ and $\psi$ such that
\begin{equation}
  \label{eqcm:33}
 \boldsymbol{\alpha}=d \chi,  \quad    \boldsymbol{\beta}= d  \psi.
\end{equation}

The form $\boldsymbol{K}$ defined by  \eqref{eqcm:15}   has the following
expression for the electromagnetic field
\begin{equation}
  \label{eqcm:16}
  \boldsymbol{K}  =2\left(\boldsymbol{F}(\alpha)
    -\frac{1}{4}\boldsymbol{\eta}|F|^2\right),
\end{equation}

We have that (see \cite{Weinstein96})
\begin{equation}
  \label{eqcm:17}
  {}^*(\boldsymbol{\eta} \wedge (\boldsymbol{F}(\alpha))=\boldsymbol{\alpha}
  \wedge 
  \boldsymbol{ \beta}.
\end{equation}
Using (\ref{eqcm:33}) we obtain 
\begin{equation}
  \label{eqcm:18}
  d (\boldsymbol{\alpha} \wedge \boldsymbol{ \beta})=0, 
\end{equation}
and hence there exist locally a 1-form $\boldsymbol{\gamma}$ such that
\begin{equation}
  \label{eqcm:19}
  d\boldsymbol{\gamma} =\boldsymbol{\alpha} \wedge
 \boldsymbol{ \beta},
\end{equation}
where $\gamma$ is given by
\begin{equation}
  \label{eqcm:20}
  \boldsymbol{\gamma}= \frac{1}{2}\left(\chi d \psi -\psi d \chi\right). 
\end{equation}
Note that
\begin{equation}
  \label{eqcm:32}
  \boldsymbol{\gamma}(\kif)=0. 
\end{equation}
From equation (\ref{eqcm:17}) we deduce
\begin{equation}
  \label{eqcm:35}
 {}^*(\boldsymbol{\eta} \wedge \boldsymbol{K})=2 \boldsymbol{\alpha} \wedge
 \boldsymbol{ \beta}.  
\end{equation}
We use the following identity valid for arbitrary 1-forms
\begin{equation}
  \label{eqcm:36}
 {}^*(\boldsymbol{\eta} \wedge \boldsymbol{K})\wedge \boldsymbol{\eta}=\eta 
{}^*\boldsymbol{K} -  {}^*\boldsymbol{\eta} ( \boldsymbol{K}(\kif)).
\end{equation}
Using (\ref{eqcm:19}) we finally obtain our main formula
\begin{equation}
 \label{eqcm:37}
 {}^*\boldsymbol{K}= 2d(\boldsymbol{\gamma} \wedge  \boldsymbol{\hat\eta}) +
 2\boldsymbol{\gamma} \wedge d \boldsymbol{\hat\eta} + {}^*\boldsymbol{\hat\eta} (
 \boldsymbol{K} \cdot \boldsymbol{\eta}), 
\end{equation}
where we have defined
\begin{equation}
  \label{eqcm:38}
\boldsymbol{\hat\eta}=  \frac{\boldsymbol{\eta}}{\eta}.  
\end{equation}
It is important to note that
\begin{equation}
  \label{eqcm:39}
  d \boldsymbol{\hat\eta}(\kif)=0.
\end{equation}
Using equation (\ref{eqcm:37}), we integrate ${}^*\boldsymbol{K}$ over a
three-surface $S$ tangential to $\eta^\mu$, with boundary $\Su$.  Using that
$\eta^\mu$ is tangential to $S$ it follows  that the restriction
of the 3-form ${}^*\boldsymbol{\hat\eta}$ to $S$ is zero.  For the second term in
(\ref{eqcm:37}) we use equations \eqref{eqcm:32} and \eqref{eqcm:39} to obtain
the same conclusion. Hence, we have
\begin{equation}
  \label{eqcm:40}
\int_S  {}^*\boldsymbol{K}= \int_S 2d(\boldsymbol{\gamma} \wedge
\boldsymbol{\hat\eta})= 2 \int_\Su \boldsymbol{\gamma} \wedge
\boldsymbol{\hat\eta},
\end{equation}
where in the last equality we have used Stokes theorem.

We summarize the previous calculation in the following lemma, which is a
re-writing of the result that  have been obtained in \cite{Simon:1984qb}.
\begin{lemma}
\label{l:ang-em}
  Consider an axially symmetric spacetime for which the Einstein-Maxwell
  equations (\ref{eqcm:2}), (\ref{eq:27}), (\ref{eqcm:51}), (\ref{eqcm:51b})
  are satisfied.  Let $S$ be an orientable three-surface, tangent to the axial
  Killing field $\eta^\mu$, with boundary (possible disconnected)
  $\Su$. Then we have
  \begin{equation}
    \label{eqcm:23}
    J(\Su)=\frac{1}{4\pi} \int_\Su \boldsymbol{\gamma} \wedge
\boldsymbol{\hat\eta},
  \end{equation}
  where $J(\Su)$ is the Komar integral given by (\ref{eqcm:1}), $
  \boldsymbol{\gamma}$ is defined in terms of the electromagnetic field by
  (\ref{eqcm:20}) and $\boldsymbol{\hat\eta}$ is given by (\ref{eqcm:38}).
\end{lemma}

We can define a `total angular momentum' which is conserved in electro-vacuum,
namely
\begin{equation}
  \label{eqcm:21}
  J_T(\Su)= \frac{1}{16\pi} \int_\Su {}^*d \boldsymbol \eta  -4\,
  \boldsymbol{\gamma} 
  \wedge \boldsymbol{\hat\eta} .
\end{equation}
We note that since the surface $\Su$ is tangent to $\eta^\mu$ and all
the fields are axially symmetric, then the surface integrals are in fact line
integrals on the quotient manifold $M\setminus SO(2)$.  

Formula (\ref{eqcm:21}) was studied in
\cite{Ashtekar:2001is}\cite{Ashtekar:2000sz} for rotating isolated horizons.
This formula  has been also recently studied at null infinity in
connection with the center of mass and general definition of angular momentum
for asymptotically flat (at null infinity) spacetimes, see \cite{kozameh11}.

A very important example where lemma \ref{l:ang-em} applies is the Kerr-Newman
black hole.  Consider the Kerr-Newman black hole with parameters $(m, a,
q)$. The total angular momentum is given by $J_T=am$. This is equal to the
Komar integral evaluated at infinity, since the electromagnetic field decay and
does not contribute at infinity.  However at the horizon the Komar angular
momentum is not $J_T$. The angular momentum at the horizon has the
decomposition (\ref{eqcm:21}).

For the Kerr-Newman black hole the Komar angular momentum at the horizon $J$
is given by (see, for example, \cite{Poisson} page 222).  
\begin{equation}
  \label{eqk:1}
  J= a \frac{r^2_+ +a^2}{2r_+} \left(1+\frac{q^2}{2a^2}\left(1- \frac{r^2_+
        +a^2}{ar_+}\right) \arctan\left(\frac{a}{r_+}\right) \right), 
\end{equation}
where $r_+$ is the horizon radius
\begin{equation}
  \label{eqk:2}
  r_+=m+(m^2-a^2-q^2 )^{1/2}.
\end{equation}
The area of the horizon is given by
\begin{equation}
  \label{eqk:3}
  A=4\pi(r^2_+ +a^2).
\end{equation}
Equation \eqref{eqk:3} is of course identical to equation \eqref{eq:11}.
As we already mentioned, it is well known that the area satisfy the inequality
\begin{equation}
  \label{eqk:4}
   A \geq  4\pi \sqrt{q^4+4J_T^2}.
\end{equation}
Which in particular implies
\begin{equation}
  \label{eqk:5}
   A \geq  8\pi |J_T|.
\end{equation}
But this inequality relates the total angular momentum $J_T$. It is a priori
not obvious if the following inequality holds
\begin{equation}
  \label{eqk:6}
   A \geq  8\pi |J|,
\end{equation}
where $J$ is given by (\ref{eqk:1}), namely the Komar integral at the horizon.
Note that this is precisely the inequality proved in theorem \ref{t:main-2} and
that the Kerr-Newman black hole satisfies all the hypothesis of that theorem.

Let us check explicitly that indeed (\ref{eqk:6}) is satisfied. The expression
(\ref{eqk:1}) is remarkably complicated, to better analyze it let us rewrite
it in the following form. Using (\ref{eqk:3}) we have
\begin{equation}
  \label{eqk:7}
  J=\frac{A}{8\pi}\epsilon,
\end{equation}
where
\begin{equation}
  \label{eqk:8}
  \epsilon=\frac{a}{r_+}\left(1+\frac{q^2}{2a^2}\left(1- \frac{r^2_+
        +a^2}{ar_+}\right) \arctan\left(\frac{a}{r_+}\right) \right), 
\end{equation}
Instead of using $a$ it is convenient to use $x=a/r_+$ as free parameter. In
terms of $x$ the function $\epsilon$ is written us
\begin{equation}
  \label{eqk:9}
  \epsilon=x\left( 1+ \frac{q^2}{2r^2_+} f(x)\right),
\end{equation}
where
\begin{equation}
  \label{eqk:10}
  f(x)=\frac{1}{x^2} \left(1- \left(\frac{1}{x}+x \right)\arctan(x)\right).
\end{equation}
We take as free parameters $(q,r_+,x)$. Note that $-1\leq x\leq 1$ and hence we have
\begin{equation}
  \label{eqk:11}
  |\epsilon| \leq  \left| 1+ \frac{q^2}{2r^2_+} f(x)\right|.
\end{equation}
Fix  $(q,r_+)$. It can be explicitly check that function $f(x)$
is non-positive and have a unique global  minimum at $x=0$ where
$f(0)=-2/3$. Hence it follows that
\begin{equation}
  \label{eqk:12}
  |\epsilon| \leq 1. 
\end{equation}

\subsection{Potentials}
\label{s:potentials}
The potentials for the axial Killing vector plays an important role in the mass
functional described in section \ref{sec:mass-axial-symmetry}. 

\begin{figure}
  \centering
   \includegraphics[width=3cm]{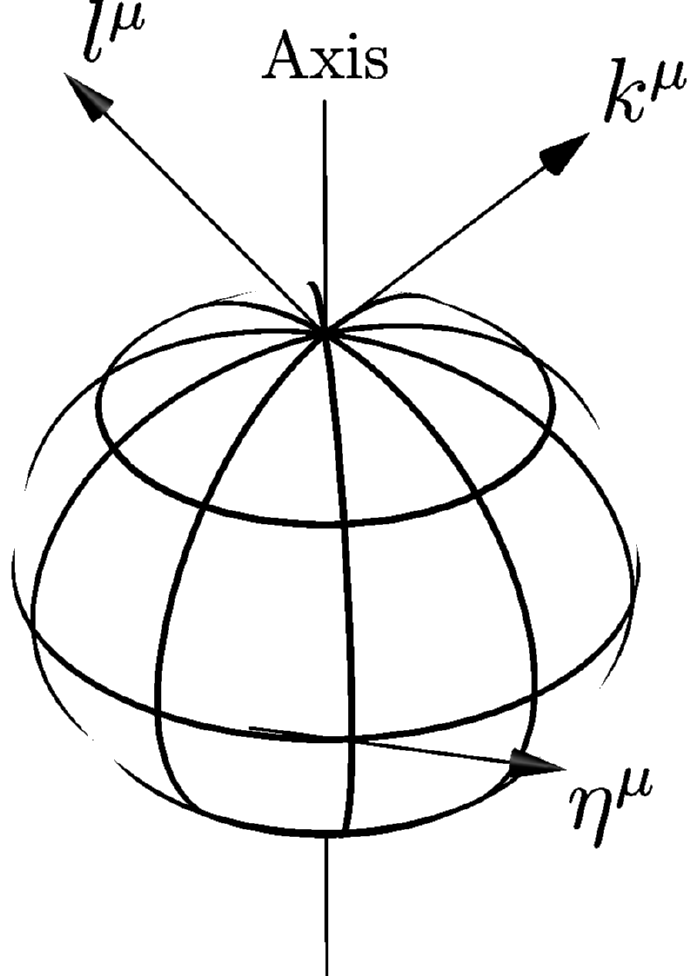}  
  \caption{Adapted tetrad for an axially symmetric two-sphere}
  \label{fig:2}
\end{figure}

It is instructive to analyze first the electric charge and its potential in
axial symmetry. Assume first that the Maxwell equations are source free. Then
we have found the potentials $\chi$ and $\psi$ defined by equation
(\ref{eqcm:33}). In particular, the potential $\psi$ determine the electric charge
over an axially symmetric two-surface $\Su$. In order to see that it is
convenient to consider a tetrad $(l^\mu,k^\mu,\xi^\mu,\eta^\mu)$ and coordinate
system $(\theta,\phi)$ adapted to an axially symmetric two-surface defined as
follows (see figure \ref{fig:2}). For simplicity we will assume that $\Su$ has
the topology of a two-sphere.  Let us consider null vectors $\ell^\mu$ and
$k^\nu$ spanning the normal plane to $\Su$ and normalized as $\ell^\mu k_\mu =
-1$, leaving a (boost) rescaling freedom $\ell'^\mu =f \ell^\mu$, $k'^\mu =
f^{-1} k^\mu$. By assumption $\eta^\mu$ is tangent to $\Su$, it has on the
surface closed integral curves and vanishes exactly at two points which are the
intersection of the axis $\Gamma$ with $\Su$. We normalize vector $\eta^\mu$ so
that its integral curves have an affine length of $2\pi$.  Let us chose a
coordinate $\phi$ on $\Su$ such that $\eta^\mu=\partial/\partial \phi$. The
other vector of the tetrad which is tangent to $\Su$ and orthogonal to
$\eta^\mu$ will be denoted by $\xi^\mu$ and assume that it has unit norm.  We
define the coordinate $\theta$ such that $\xi^\mu$ is proportional to
$\partial/\partial \theta$ and such that $\theta=\pi,0$ are the poles of $\Su$.

The induced metric and the volume element on $\Su$ (written as spacetime
projectors) are given by $q_{\mu\nu}=g_{\mu\nu}+\ell_\mu k_\nu+\ell_\nu k_\mu$
and $\epsilon_{\mu\nu}=2^{-1}\epsilon_{\lambda\gamma\mu\nu}\ell^\lambda k^\gamma$
respectively. The area measure on $\Su$ is denoted by $\ds$.  Since the surface
is axially symmetric we have ${\cal L}_\eta q_{\mu\nu}=0$.

Using this tetrad, the charge over an axially symmetric surface $\Su$ is
written as follows 
\begin{equation}
  \label{eq:37}
  q(\Su)=\frac{1}{2\pi}\int_\Su \eta^{-1/2} \xi^\mu\beta_\mu \, \ds. 
\end{equation}
where $\beta_\mu$ is defined by (\ref{eqcm:30}). In the source free case we can
use the potential $\psi$ defined by (\ref{eqcm:33}) to obtain  
\begin{equation}
  \label{eq:36}
  q(\Su)=\frac{1}{2\pi} \int_\Su \eta^{-1/2} \xi^\mu \nabla_\mu \psi \, \ds =
  \int_0^\pi \partial_\theta \psi \, d\theta.
=\psi(\pi)-\psi(0),
\end{equation}
That is, the charge is given by the difference of the value of the potential
$\psi$ at the poles of the surface $\Su$.

We have seen that the potential $\psi$ is only defined in the source-free case
as scalar function in the spacetime. However, on the surface $\Su$ it is
always possible to define a potential $\bar \psi$ by the equation
\begin{equation}
  \label{eq:39}
  \xi^\mu\psi_\mu = \xi^\mu \nabla_\mu \bar \psi
\end{equation}
on the surface, since this equation involves only a derivative with respect to
$\theta$. The potential $\bar\psi$ is only defined on the surface. By
definition, in the source-free case where the potential $\psi$ also exists, the
two functions are equal up to a constants on the surface. This constant is
irrelevant since it does not affect the charge.


Consider now the potential for the angular momentum. 
The twist vector of $\eta^\mu$ is defined by
\begin{equation}
  \label{eqt:2}
  \omega_\mu= \epsilon_{\mu\nu\lambda \gamma }\eta^\nu \nabla^\lambda\eta^\gamma. 
\end{equation}
It is a well known result that the vacuum equations $R_{\mu\nu}=0$ imply that
\begin{equation}
  \label{eqt:3}
  d \boldsymbol{\omega}=0,
\end{equation}
where $\boldsymbol{\omega}$ is the 1-form defined by (\ref{eqt:2}).  Hence
there exist a potential $\omega$ such that
\begin{equation}
  \label{eq:31}
   \boldsymbol{\omega}=d \omega.
\end{equation}
The function $\omega$ is the twist potential of the Killing vector $\eta^\mu$,
it contains all the information of the angular momentum as we will see.  In the
non-vacuum case, the twist potential is not defined.  In the following we will
not assume vacuum.

In terms of the adapted tetrad we have that the Komar expression is given by
\begin{equation}
  \label{eqt:8}
  J=\frac{1}{8\pi}\int_{\cal S} \nabla^{\mu} \eta^\nu l_{\mu}k_\nu \, \ds.
\end{equation}
We  the relation 
\begin{equation}
  \label{eqt:4}
  \nabla_\mu \eta_\nu=\frac{1}{2}\eta^{-1}  \epsilon_{\mu\nu \lambda \gamma} 
\eta^\lambda \omega^\gamma+ \eta^{-1}  \eta_{[\nu} \nabla_{\mu]}\eta. 
\end{equation}
to write the Komar integral in the following form
\begin{equation}
  \label{eqt:5}
   J=\frac{1}{16\pi}\int_{\cal S}  \eta^{-1}  \epsilon_{\mu\nu\lambda \gamma } \eta^\lambda
   \omega^\gamma  \ell^{\mu} k^\nu \, \ds. 
\end{equation}
It is clear that the vector defined by 
\begin{equation}
  \label{eqt:6}
  \xi_\gamma=  \eta^{-1/2}  \epsilon_{\mu\nu\lambda \gamma } \eta^\lambda \ell^{\mu} k^\nu.   
\end{equation}
is orthogonal to $l^\mu,k^\mu,\eta^\mu$ and it has unit norm and hence is the
other member of the tetrad.  Then we have
\begin{equation}
  \label{eqtc:7}
   J=\frac{1}{16\pi}\int_{\cal S}  \eta^{-1/2}  \xi^\mu \omega_\mu \, \ds.
\end{equation}
We emphasize that this expression is valid only for axially symmetric surfaces.

If we assume vacuum, then we can use the twist potential defined by
(\ref{eq:31}) to obtain 
\begin{equation}
  \label{eqt:7}
   J=\frac{1}{8}\int_{0}^\pi  \partial_\theta\omega \, d\theta
   =\frac{1}{8} \left(\omega(\pi)-\omega(0) \right).
\end{equation}
Equations (\ref{eqtc:7}) and (\ref{eqt:7}) are the analog to equations
(\ref{eq:37}) and (\ref{eq:36}) for the charge.

We consider now the non-vacuum case. We define the following vector (a part of
the extrinsic curvature of $\Su$ with the role of a connection on its normal
cotangent bundle, see e.g. the discussion in \cite{Gou05} \cite{Booth:2006bn})
\begin{equation}
  \label{eqt:10}
  \Omega^{(\ell)}_\mu = -k^\gamma {q^\lambda}_\mu \nabla_\lambda \ell_\gamma \ .
\end{equation}
Since $\Su$ is axially symmetric, the tetrad can be chosen such that 
\begin{equation}
  \label{eqt:11}
  \pounds_\eta l^\mu=\pounds_\eta k^\mu=0.
\end{equation}
In particular this implies that 
\begin{equation}
  \label{eqt:12}
  l^\mu\nabla_\mu\eta^\nu- \eta^\mu\nabla_\mu l^\nu=0.
\end{equation}
Using this equation we obtain that
\begin{equation}
  \label{eqt:13}
   \eta^\mu \Omega^{(\ell)}_\mu = -k^\mu \eta^\nu \nabla_\nu \ell_\mu=-k^\mu
   \ell^\nu \nabla_\nu \eta_\mu. 
\end{equation}
From this equation we  get  the following  equivalent expression for the Komar angular
momentum (see, for example, \cite{jaramillo11c})
\begin{equation}
  \label{eqt:14}
  J=\frac{1}{8\pi}\int_{\cal S}  \eta^\mu \Omega^{(\ell)}_\mu \, \ds. 
\end{equation}
Remarkably, as we will see in the following, the vector $\Omega^{(\ell)}_\mu$
determines even in the non-vacuum case a potential on the surface (we will
denote it by $\bar \omega$) which coincide with the twist potential $\omega$
defined above in the vacuum case.  We emphasize that the potential $\bar\omega$
will be only defined on the two-surface $\Su$, in contrast to the twist
potential $\omega$ which (in the vacuum case) is defined by equations
(\ref{eqt:3}) and (\ref{eq:31}) on a region of the spacetime.

By construction, the vector $\Omega^{(\ell)}_\mu$ is tangent to the surface
$\Su$. Since we have assumed that $\Su$ has the $\mathbb{S}^2$ topology, there
exists functions $\hat \omega$ and $\lambda$ such that the vector has the
following decomposition on $\Su$  in terms of a divergence-free and an exact form
\begin{equation}
  \label{eq:73}
  \Omega^{(\ell)}_A=T_A+\nabla_A\lambda, \quad T_A=\epsilon_{AB}\nabla^B\hat \omega.
\end{equation}
The functions $\hat \omega$ and $\lambda$ are fixed up to a constant. In
equation (\ref{eq:73}) we have used the capital indices (which run from $1$ to
$2$) to emphasizes that this is an intrinsic equation on the two-surface $\Su$.

By assumption both $\Omega^{(\ell)}_A$ and the intrinsic metric on $\Su$ are
axially symmetric, then it follows that the functions $\hat \omega$ and
$\lambda$ are also axially symmetric (i.e. they depend only on $\theta$). In
particular it follows that
\begin{equation}
  \label{eq:76}
  \xi^AT_A=0, \quad \eta^A\nabla_A \lambda=0,
\end{equation}
where $\xi^A$ is the vector tangent to $\Su$ previously defined. Since the norm
$\eta$ is also an axially symmetric function, equation (\ref{eq:76}) implies that
\begin{equation}
  \label{eq:77}
 T^A D_A\eta =0.
\end{equation}
Equation (\ref{eq:77}) ensures the integrability conditions for the  existence
of  a function $\bar\omega$ such that
\begin{equation}
  \label{eq:78}
  T_A=\frac{1}{2\eta}\epsilon_{AB}\nabla^B\bar \omega.
\end{equation}
Note that equation (\ref{eq:78}) is valid only in axial symmetry. Collecting
these results we obtain the decomposition
\begin{equation}
  \label{eqt:15}
  \Omega^{(\ell)}_A=  \frac{1}{2\eta} \epsilon_{A B}\nabla^B \bar \omega
  +\nabla_A \lambda. 
\end{equation}
In particular, using (\ref{eq:76}),  we obtain
\begin{equation}
  \label{eq:79}
\eta^A  \Omega^{(\ell)}_A= \frac{1}{2\eta} \epsilon_{A B}\eta^A \nabla^B \bar \omega.
\end{equation}
We write equation (\ref{eq:79}) with spacetime indices and we use the
 following representation for the tetrad vector $\xi^\mu$ 
\begin{equation}
  \label{eqt:17}
  \xi_\nu=\eta^{-1/2} \epsilon_{\mu\nu }\eta^\mu,
\end{equation}
to finally obtain
\begin{equation}
  \label{eqt:18}
  \eta^\mu \Omega^{(\ell)}_\mu =  \frac{1}{2}\eta^{-1/2}\xi^\mu \nabla_\mu\bar \omega.
\end{equation}
Using equation (\ref{eqt:14}) an integrating we finally get 
\begin{equation}
  \label{eqt:21}
  J=\frac{1}{8}\int_{0}^\pi  \partial_\theta\bar \omega \, d\theta
   =\frac{1}{8} \left(\bar \omega(\pi)-\bar \omega(0) \right). 
\end{equation}
The relevance of this construction is that the function $\bar \omega$, which is
only defined at the surface $\Su$, plays the role of a potential that can be
defined in the non-vacuum case \cite{Jaramillo:2011pg}. To see the relation between $\bar \omega$ and
$\omega$, note that we  have
\begin{equation}
  \label{eqt:19}
  \xi^\mu\omega_\mu=\xi^\mu \nabla_\mu \bar \omega.
\end{equation}
This equation is valid always (i.e. non-vacuum) and it gives the function
$\bar\omega$ in terms of the twist vector $\omega_\mu$. In the non-vacuum case the
twist potential $\omega$ is not defined but the function $\bar \omega $ is always well
defined. In fact, we can take (\ref{eqt:19}) as definition of $\bar \omega$,
since the right hand side is only a derivative with respect to $\theta$. 
On the other hand, in the vacuum case  equation (\ref{eqt:19}) implies that
$\omega$ and $\bar \omega$ differs by a constant which is irrelevant since it
does not contribute to the angular momentum.

\section{Mass in axial symmetry}
\label{sec:mass-axial-symmetry}
The main goal of this section is to present the mass formula for axially
symmetric data (\ref{eqq:103}) and the mass functional for two-surfaces
(\ref{eqq:44}). We also discuss their geometrical properties in connection with
harmonic maps.

\subsection{Axially symmetric initial data}
\label{sec:axially-symm-init}
The global geometrical inequality (\ref{eq:42}) is studied on axially
symmetric, asymptotically flat initial data set. In order to present the mass
formula, we first review the basic definitions and properties of this kind of
initial data.  For simplicity we concentrate only in the pure vacuum case (see
\cite{Chrusciel:2009ki} \cite{Costa:2009hn} for the electro-vacuum case).

An \emph{initial data set} for the Einstein vacuum equations is given
by a triplet $(S, h_{ij}, K_{ij})$ where $S$ is a connected
three-dimensional manifold, $ h_{ij} $ a Riemannian metric, and $ K_{ij}$ a
symmetric tensor field on $S$. 
The fields are assumed to satisfy  the vacuum constraint equations
\begin{align}
 \label{const1}
   D_j   K^{ij} -  D^i   K= 0,\\
 \label{const2}
   R - K_{ij}   K^{ij}+  K^2=0, 
\end{align}
where $ {D}$ and $ R$ are the Levi-Civita connection and
the scalar curvature associated with $ {h}_{ij}$, and $ K = K_{ij} h^{ij}$. In
these equations the indices are moved with the metric $ h_{ij}$ and its inverse
$ h^{ij}$. 

The initial data is called \emph{maximal} if
\begin{equation}
  \label{eq:48}
  K=0.
\end{equation}
This condition is crucial because it implies via the Hamiltonian constrain
(\ref{const2}) that the scalar curvature $R$ is non-negative. In fact on the
right hand side of equation (\ref{const2}) it is possible to add a non-negative
matter density and all the arguments behind the proof of theorem \ref{t:main-1}
will also apply since they rely on lower bounds for $R$.

The manifold $S$ is called \emph{Euclidean at infinity}, if there exists a
compact subset $\mathcal{K}$ of $S$ such that $S\setminus \mathcal{K}$ is the
disjoint union of a finite number of open sets $U_k$, and each $U_k$ is
diffeomorphic to the exterior of a ball in $\Rt$. Each open set $U_k$ is called an
\emph{end} of $S$. Consider one end $U$ and the canonical coordinates $x^i$ in
$\Rt$ which contains the exterior of the ball to which $U$ is diffeomorphic
to. Set $r=\left( \sum (x^i)^2 \right)^{1/2}$. An initial data set is called
\emph{asymptotically flat} if $S$ is Euclidean at infinity, the metric $
h_{ij}$ tends to the euclidean metric and $ K_{ij}$ tends to zero as $r\to
\infty$ in an appropriate way.  These fall off conditions (see \cite{Bartnik86}
\cite{chrusciel86} for the optimal fall off rates) imply the existence of the
total mass $m$ (or ADM mass \cite{Arnowitt62}) defined at each end $U$ by
\begin{equation}
  \label{eq:30adm}
m=\frac{1}{16\pi}\lim_{r\to \infty} \int_{\Su_r} \left(\partial_j
    h_{ij}-\partial_i 
    h_{jj}\right ) s^i \ds,  
\end{equation}
where $\partial$ denotes partial derivatives with respect to $x^i$,
$\Su_r$ is the euclidean sphere $r=constant$ in $U$, $s^i$ is its
exterior unit normal and $\ds$ is the surface element with respect to the
euclidean metric. 

The angular momentum is also defined as a surface integral at infinity
(supplementary fall off conditions must be imposed, see, for example, the
review articles \cite{Szabados04}, \cite{jaramillo11c} and reference
therein). Let $\beta^i$ be an infinitesimal generator for rotations with
respect to the flat metric associated with the end $U$, then the angular
momentum $J$ in the direction of $\beta^i$ is given by
\begin{equation}
  \label{eq:35}
  J=\frac{1}{8\pi} \lim_{r\to \infty} \int_{\Su_r} (K_{ij}
  -Kh_{ij})\beta^i s^j  \ds.
\end{equation}

The above discussion applies to general asymptotically flat initial data. We
have seen that at any end the total mass $m$ and the total angular momentum $J$
are well defined quantities. We consider now axially symmetric initial data. 
In analog way as the spacetime definition \ref{d:ax}, we say that the
Riemannian manifold $(S,h_{ij})$ is axially symmetric if its group of
isometries has a subgroup isomorphic to $SO(2)$.  We will denote by $\eta^i$
the Killing field generator of the axial symmetry and by $\Gamma$ the axis. The
initial data set is called axially symmetric if in addition $\eta^i$ is also a
symmetry of $K_{ij}$, namely
\begin{equation}
  \label{eq:33}
  \pounds_\eta K_{ij}  =0.
\end{equation}
On an axially symmetric spacetime there exist axially symmetric initial
condition, conversely axially symmetric initial data evolve into an axially
symmetric spacetime. However, on an axially symmetric spacetime it is also
possible to take initial conditions which are not axially symmetric in the
sense defined above. These conditions will of course also evolve into an
axially symmetric spacetime, but the Killing vector is `hidden' on them (these
kind of initial data were studied in \cite{beig97}). We will not consider this
kind of data here since on an axially symmetric spacetime it is always possible
to chose initial conditions which are explicitly axially symmetric in the sense
defined above.

For axially symmetric data we have  the Komar integral discussed in
section \ref{sec:angul-moment-axial}.  This integral can be calculated in terms
on the initial data if we chose $\Su\subset S$. 
There exist a very simple relation expression for the Komar integral
on an initial data, namely
\begin{equation}
  \label{eq:38}
  J=\frac{1}{8\pi}\int_\Su K_{ij} \eta^i s^j \ds.
\end{equation}
The equivalence between (\ref{eq:38}) and the original definition
(\ref{eqcm:1}) can be seen as follows. Consider the tetrad adapted to $\Su$
defined in section \ref{s:potentials}. Assume that $\Su\subset S$. By
assumption, the axial Killing vector is tangent to the three-dimensional
surface $S$. Denote by $n^\mu$ the timelike unit normal to the three-surface
$S$, and let $s^\mu$ be the spacelike unit normal to the two-surface $\Su$
which lies on $S$. These vectors can be written in term of $\ell^\mu$ and
$k^\mu$ as follows 
\begin{equation}
  \label{eq:80}
  n^\mu= \frac{1}{\sqrt{2}} (\ell^\mu+k^\mu ),\quad  s^\mu= \frac{1}{\sqrt{2}} (\ell^\mu-k^\mu ), 
\end{equation}
Using these expression, we can write the integrand in 
(\ref{eqcm:1}) as
\begin{equation}
  \label{eq:81}
   J(\Su)=\frac{1}{8\pi}\int_\Su  n^\lambda s^\gamma  \nabla_\lambda \eta_\gamma.
\end{equation}
The second fundamental form can be written (as spacetime tensor) in terms of
the unit normal $n^\mu$ as
\begin{equation}
  \label{eq:82}
  K_{\mu\nu}=-h^\lambda_\mu \nabla_\lambda n_\nu.
\end{equation}
Using that $\eta^\mu n_\mu=0$ and that $s^\mu$ is tangent to $S$, from
(\ref{eq:82}) we obtain
\begin{equation}
  \label{eq:84}
 K_{\mu\nu}s^\mu \eta^\nu=  n^\nu s^\lambda \nabla_\lambda \eta_\nu. 
\end{equation}
From (\ref{eq:84}) and (\ref{eq:81}) we obtain (\ref{eq:38}).

Comparing expression (\ref{eq:38}) with (\ref{eq:35}) we see that if we chose
in (\ref{eq:35}) the vector $\beta^i=\eta^i$ (that it, the generator of axial
rotations) then these to expression are equivalent since $\eta^is_i=0$ for an
sphere at infinity.

It is possible to calculate the potential $\omega$ for the spacetime Killing
field defined in section \ref{sec:angul-moment-axial} in terms of $K_{ij}$ as
follows.  Define the vector $K_i$ by
\begin{equation}
  \label{eq:40}
  K_i=\epsilon_{ijk}S^j \eta^k, \quad S_i=K_{ij}\eta^j,
\end{equation}
where $\epsilon_{ijk}$ is the volume element with respect to the metric
$h_{ij}$. 
Then, as a consequence of equations (\ref{const1}) and (\ref{eq:33}) we have that
\begin{equation}
  \label{eq:41}
  D_{[i} K_{j]}=0.
\end{equation}
Then the potential $\omega$ is defined by
\begin{equation}
  \label{eq:44}
   K_i=-\frac{1}{2} D_i \omega.
\end{equation}
It can be checked that this is the same potential as defined in the previous
section (see \cite{Dain:2008xr}) evaluated on the initial surface. The
importance of the expression (\ref{eq:44}) is that allow to calculate $\omega$
in terms of the initial conditions.
 
The functions $\omega$ and $\eta$ have the information of the stationary part
of the initial conditions. The dynamical part is characterized by the functions
$\eta'$ and $\omega'$ defined by 
\begin{equation}
  \label{eq:45}
\eta'= -2K_{ij}\eta^i\eta^j, \quad  \omega'=\epsilon_{ijk}\eta^i  D^j \eta^k.  
\end{equation}
The notation comes from the fact that they are related with the time derivative
of this functions, namely if $n^\mu$ is the timelike normal of the initial
data, we have (see \cite{Dain:2008xr})
\begin{equation}
  \label{eq:46}
\eta'= n^\mu\nabla_\mu \eta, \quad   \omega'= n^\mu\nabla_\mu \omega.  
\end{equation}
The whole initial data can be constructed in terms of $(\eta, \omega; \eta',
\omega')$. These functions are the initial data for the wave map equations that
characterize the evolution in axial symmetry.  This is clearly seen in the
quotient representation (see \cite{dain10}). For our present purpose the
relevant part of the initial conditions is contained in $(\eta,\omega)$.

\begin{figure}
  \centering
   \includegraphics[width=5cm]{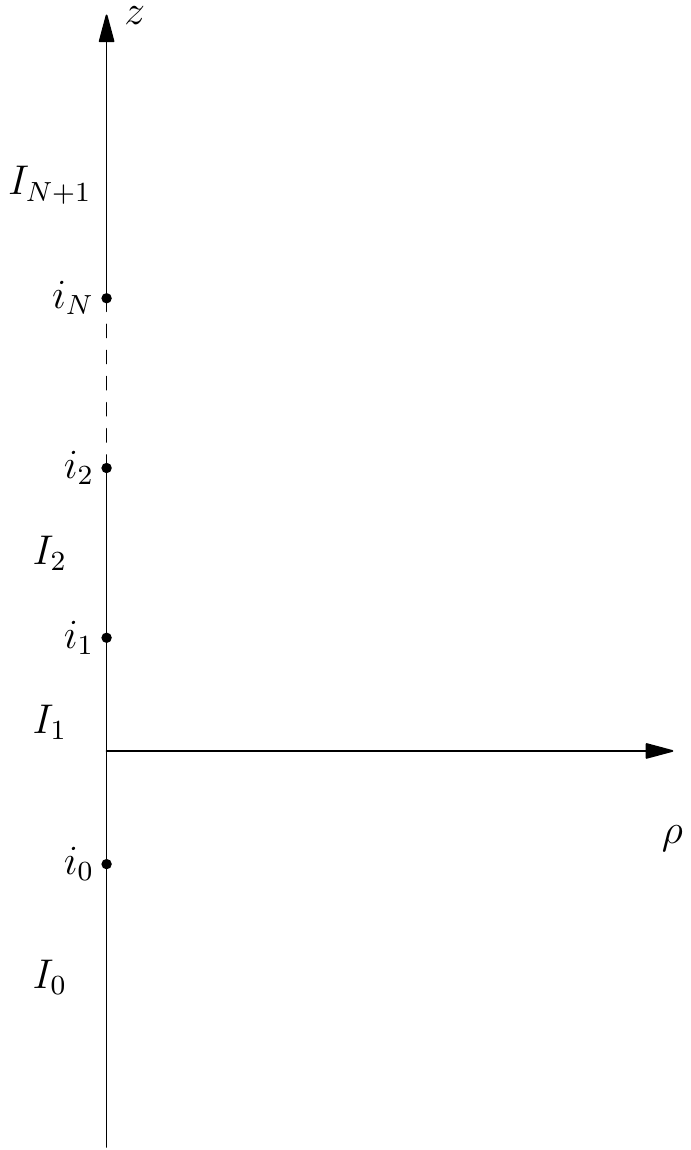}  
  \caption{Axially symmetric data with $N$ asymptotic ends.}
  \label{fig:5}
\end{figure}

So far we have discussed local implications of axial symmetry. Suppose that $S$
is simply connected and asymptotically flat (with, possible, multiple ends).
It can be proved (see \cite{Chrusciel:2007dd}) that in such case the analysis
essentially reduces to consider manifold of the form $S=\Rt\setminus
\sum_{k=0}^{N} i_k $ where $i_k$ are points in $\Rt$.  These points represent the
extra asymptotic ends of $S$. 
Moreover, in \cite{Chrusciel:2007dd} it has been proved that on $S$ there exists the
following global coordinate systems which will be essential in what follows.
\begin{lemma}[Isothermal coordinates]
  \label{isothermal}
  Consider an axially symmetric, asymptotically flat (with possible multiple
  ends) Riemannian manifold $(S, h_{ij})$, where $S$ is assumed to be simply
  connected. Then, there exist a global coordinate system system
  $(\rho,z,\varphi)$ such the metric has the following form
\begin{equation}
  \label{eq:44b}
   h=  e^{(\sigma+2q)}(d\rho^2+dz^2)+ \rho^2 e^\sigma  (d\varphi + A_\rho
 d\rho+ A_z dz)^2,
\end{equation}
where the functions $\sigma,q, A_\rho, A_z$ do not depend on $\varphi$.  In
these coordinates, the axial Killing vector is given by $\eta=
\partial/\partial \varphi$ and the square of its norm is given by 
\begin{equation}
  \label{eq:83}
\eta=\rho^2e^{\sigma}.
\end{equation} 
\end{lemma}
Using this coordinates system, the end points $i_k$ are located at the axis
$\rho=0$ of $\Rt$.  Define the intervals $I_k$, $1\leq k\leq N-1 $, to be the
open sets in the axis between $i_k$ and $i_{k-1}$, we also define $I_0$ and
$I_N$ as $z< i_0$ and $z >i_N$ respectively (see figure \ref{fig:5}).  The
manifold $S$ is Euclidean at infinity with $N+1$ ends.  In effect, for each
$i_k$ take a small open ball $B_k$ of radius $r_{(k)}$, centered at $i_k$,
where $r_{(k)}$ is small enough such that $B_k$ does not contain any other
$i_{k'}$ with $k'\neq k$. Let $\bar B_R$ be a closed ball, with large radius
$R$, such that $\bar B_R$ contains all points $i_k$ in its interior. The
compact set $\mathcal{K}$ is given by $\mathcal{K}= \bar B_R \setminus
\sum_{k=1}^N B_k$ and the open sets $U_k$ are given by $B_k\setminus i_k$, for
$1 \leq k \leq N$, and $U_0$ is given by $\Rt \setminus \bar B_R$.  Our choice of
coordinate makes an artificial distinction between the end $U_0$ (which
represent $r\rightarrow \infty$) and the other ones.  This is convenient for
our purpose because we want to work always at one fixed end.
We emphasize that the isothermal coordinates $(\rho,z,\varphi)$ (and hence the
corresponding spherical radius $r=\sqrt{\rho^2+z^2}$) are globally defined on
$S$. In what follows we will  always use these coordinates. 

The smoothness of the initial data on the axis implies that the potential
$\omega$ is constant on each interval $I_k$. These constants are directly
related with the angular momentum of the end points $i_k$. In effect, 
the angular momentum of an end $i_k$ is defined to be the Komar integral with
respect to a surface $\Su_k$  that encloses only that point. Using the formula
(\ref{eqt:7}) we obtain
\begin{equation}
  \label{eq:45gg}
J_k\equiv J(\Su_k)=\frac{1}{8}\left (\omega|_{I_k} -\omega|_{I_{k-1}}\right ).
\end{equation}
Let $\Su_0$ a surface that enclose all the end points $i_k$, then the total
angular momentum of the end $r\to \infty$ is given by
\begin{equation}
  \label{eq:46gg}
J_0\equiv J(\Su_0)=\frac{1}{8}\left (\omega|_{I_0} -\omega|_{I_{N}}\right ),
\end{equation}
which is equivalent to 
\begin{equation}
  \label{eq:47gg}
  J_0=\sum_{k=1}^N J_k. 
\end{equation}

The mass (\ref{eq:30adm}) is defined as a boundary integral at infinity. For
axially symmetric initial data there is an equivalent representation of the
mass as a positive definite integral over the three-dimensional slice. This
formula was discovered by Brill \cite{Brill59} and allow him to prove the first
version of the positive mass for pure vacuum axially symmetric gravitational
waves. This formula was generalized to include metrics with non hypersurface
orthogonal Killing vectors (i.e.  with non-zero $A_\rho$ and $A_z$ in the
representation (\ref{eq:44b})) in \cite{Gibbons06} and for non-trivial
topologies in \cite{Dain05c}.  In particular, this includes the topology of the
Kerr initial data.  See also \cite{Dain:2008xr} and \cite{dain10} for related
discussion on this mass formula.

Take isothermal coordinates and assume that the initial data are maximal
(i.e. condition (\ref{eq:48}) holds).  Then the total mass (\ref{eq:30adm}) is
given by the following positive definite integral
\begin{equation}
  \label{eqq:15}
m= \frac{1}{32\pi}\int_{\Rt}   \left[|\partial \sigma |^2
+ \frac{e^{(\sigma+2q)} \omega'^2}{\eta^2}  + 2e^{(\sigma+2q)} K^{ij}K_{ij}
\right ] \, \dv, 
\end{equation}
where $|\partial \sigma |^2= \sigma^2_{,\rho} +\sigma^2_{,z}$ and $\dv$ is the
flat volume element in $\Rt$, namely $\dv=\rho dz d\rho d\varphi$. Since the
functions are axially symmetric the integration in $\varphi$ is trivial, we
keep however this notation because it will be important in section
\ref{sec:harmonic-maps}. 

The remarkable fact of the integral (\ref{eqq:15}) is that it applies also to
the case of multiples ends, the integral is taking over all the extra ends
$i_k$ since it cover the whole $\Rt$.

The integrand in (\ref{eqq:15}) can be further decomposed as follows (see
\cite{dain10} for details)
\begin{equation}
  \label{eqq:103}
  m= \frac{1}{32\pi}\int_{\Rt} \left[  \frac{e^{(\sigma+2q)}}{\eta^2}\left(\eta'^2 + 
\omega'^2 \right)+ F^2  +  |\partial
 \sigma|^2 + \frac{|\partial \omega|^2}{\eta^2}  \right] \, \dv .    
\end{equation}
Where the term $F^2$ is essentially the square of the second fundamental form of
the two-surface that is obtained by the quotient of the manifold $S$ by the
symmetry $SO(2)$ (this two-surface is a half plane $\rho\geq 0$ minus the end
points $i_k$). The important point in this representation is that it clearly
separates the dynamical and the stationary terms in the mass formula. In fact,
this formula can be seen as the geometrical conserved energy of the wave map
that appear in the evolution equations (see \cite{dain10}).  
 
The first three terms in the integrand of (\ref{eqq:103}) correspond to the
dynamical part for the mass. They vanished for stationary solutions. The last
two terms correspond to the stationary part. They give the total mass for Kerr.
Explicitly the stationary part is given by
\begin{equation}
  \label{eqq:5c}
 \mf(\sigma,\omega)= \frac{1}{32\pi}\int_{\mathbb{R}^3}
  \left(|\partial  \sigma |^2  + \rho^{-4}e^{-2\sigma} |\partial \omega |^2
  \right) \dv, 
\end{equation}
An hence we obtain the important bound
\begin{equation}
  \label{eq:47}
  m\geq  \mf(\sigma,\omega).
\end{equation}

We have presented the mass formula but we have not discussed  the asymptotic
conditions that ensures that this integral is well defined. We discuss them now
trying to focus only in the essential aspects, for the technical details we
refer to \cite{Dain06c} and \cite{Chrusciel:2007dd}.  All the important
features can be seen already in the term $|\partial \sigma|^2$ in
(\ref{eqq:5c}). We will concentrate in this term only in what follows.

It is instructive to see how the function $\sigma$ behaves for the 
Kerr black hole initial data with mass $m$.  These
initial data have two asymptotic ends, namely we have only one extra end point
$i_1$. In the limit $r\to \infty$ (that is, the end $U_0$) we have
\begin{equation}
  \label{eq:70}
  \sigma = \frac{2m}{r}+O(r^{-2}). 
\end{equation}
At the end $i_1$ we have
\begin{equation}
  \label{eq:74}
  \sigma = -4\log(r)+O(1)
\end{equation}
for the non-extreme case, and
\begin{equation}
  \label{eq:75}
   \sigma = -2\log(r)+O(1),
\end{equation}
for the extreme case. We see that the fall off behavior at the end $i_1$ is
different for the non-extreme and extreme case, this reflect of course the
difference between an asymptotically flat end and a cylindrical end (see
figures \ref{fig:4} and \ref{fig:3} respectively).

To include this kind of fall off behavior let us consider the following
conditions. In the limit   $r\to \infty$ we impose 
\begin{equation}
  \label{eq:69}
  \sigma=o(r^{-1/2}), \quad \partial \sigma=o(r^{-3/2}).
\end{equation}
At the end point $i_k$ we impose
\begin{equation}
  \label{eq:71}
\sigma   =o(r_{(k)}^{-1/2}), \quad \partial \sigma=o(r_{(k)}^{-3/2}),  
\end{equation}
Despite the formal similarity of (\ref{eq:71}) and (\ref{eq:69}) they are of a
very different nature. Conditions (\ref{eq:69}) are essentially the standard
asymptotic flat conditions on the metric, they include in particular the
behavior (\ref{eq:70}) of the Kerr initial data. On the other hand, conditions
(\ref{eq:71}) on the extra ends $i_k$ are much more general than asymptotic
flatness. Note that asymptotic flatness at the ends $i_k$ in these coordinates
essentially implies a behavior like (\ref{eq:69}). But conditions (\ref{eq:74})
includes also (\ref{eq:75}) (in fact also any logarithmic behavior with any
coefficient). 

The motivation for the fall off conditions  (\ref{eq:71}) and (\ref{eq:69}) is
that they ensures that the integral of $|\partial \sigma|^2$ is finite both at
infinity and at the end points $i_k$. And the advantage of condition
(\ref{eq:71}) is that it  encompasses both kind of behavior: asymptotically flat and
cylindrical. This is the main motivation of the Brill data definition that
appears in theorem \ref{t2} (see  \cite{Dain06c} for details).

\subsection{Harmonic maps}
\label{sec:harmonic-maps}

The crucial property of the mass functional defined in \eqref{eqq:5c}
is its relation to the energy of harmonic maps from $\Rt$ to the
hyperbolic plane $\mathbb{H}^2$: they differ by a boundary term. To see this
relation consider first the mass functional $\mf$ defined on a bounded region
$\Omega$ such that $\Omega$ does not intersect the axis $\Gamma$ (given by
$\rho=0$) 
\begin{equation}
 \label{eqq:3}
 \mf_{\Omega}= \frac{1}{32\pi}\int_{\Omega}
  \left(|\partial  \sigma |^2  + \rho^{-4}e^{-2\sigma} |\partial \omega |^2
  \right)    \dv.  
\end{equation}
We consider this integral for general functions $\sigma$ and $\omega$ which are
not necessarily axially symmetric. 
The function $\log \rho$ is harmonic outside the axis, that is
\begin{equation}
  \label{eq:72}
  \Delta \log \rho=0, \text{ in } \Rt\setminus\Gamma,
\end{equation}
where $\Delta$ is the flat Laplacian in $\Rt$. Using equation (\ref{eq:72}) we
obtain the following identity
\begin{equation}
  \label{eqq:18}
 \mf_{\Omega}= \mf'_{\Omega}- \oint_{\partial \Omega}  4\frac{\partial
  \log\rho}{\partial s} (\log\rho+\sigma)\, \ds,  
\end{equation}
where the derivative is taken in the normal direction to the boundary $\partial
\Omega$  and  $\mf'_{\Omega}$ is given by
\begin{equation} 
  \label{eqq:19}
\mf'_{\Omega}=\frac{1}{32\pi}\int_{\Omega}
  \left(\frac{ |\partial  \eta |^2  +   |\partial \omega |^2}{\eta^2}
  \right) \dv. 
\end{equation}
Recall that $\eta$ is defined  by (\ref{eq:83}).

The functional $\mf'_{\Omega}$ defines an energy for  maps
$(\eta,\omega):\Rt \to \mathbb{H}^2$ where $\mathbb{H}^2$ denotes the
hyperbolic plane $\{(\eta,\omega): \eta>0\}$, equipped with the negative constant
curvature metric
\begin{equation}
  \label{eqq:53}
ds^2= \frac{d\eta^2+d\omega^2}{\eta^2}. 
\end{equation}
The Euler-Lagrange equations for the energy $\mf'_{\Omega}$ are given by
\begin{align}
  \label{eqq:ha1}
  \Delta \log \eta &= -\frac{|\partial \omega|^2}{\eta^2},\\
\label{eqq:ha2}
\Delta \omega   & =
2 \frac{\partial \omega\partial \eta}{\eta}. 
\end{align}
The solutions of
\eqref{eqq:ha1}--\eqref{eqq:ha2}, i.e, the critical points of
$\mf'_{\Omega}$, are called harmonic maps from $\Rt \to \mathbb{H}^2$.
Since $\mf_{\Omega}$ and $\mf'_{\Omega}$ differ only by a boundary
term, they have the same Euler-Lagrange equations. 

Harmonic maps have been intensively studied, in particular the Dirichlet
problem for target manifolds with negative curvature has been solved
\cite{Hamilton75} \cite{Schoen83} \cite{Schoen82}, and
\cite{Hildebrandt77}. The last article is particularly relevant here.  However,
these results do not directly apply in our case because the equations are
singular at the axis. One of the main technical complications in the proof of
theorem \ref{t:main-1} and \ref{t2} is precisely how to handle the singular
behavior at the axis. 

We present in what follows the main strategy in the proof of the inequality
(\ref{eq:10}) (we follow the approach presented in \cite{Chrusciel:2007ak} and
\cite{Costa:2009hn}).  The core of the proof is the use of a theorem due to
Hildebrandt, Kaul and Widman \cite{Hildebrandt77} for harmonic maps. In that
work it is shown that if the domain for the map is compact, connected, with
nonvoid boundary and the target manifold has negative sectional curvature, then
minimizers of the harmonic energy with Dirichlet boundary conditions exist, are
smooth, and satisfy the associated Euler-Lagrange equations. That is, harmonic
maps are minimizers of the harmonic energy for given Dirichlet boundary
conditions. Also, solutions of the Dirichlet boundary value problem are unique
when the target manifold has negative sectional curvature. Therefore, we want
to use the relation between $\mf$ and the harmonic energy $\mf'$ in order to
prove that minimizers of $\mf'$ are also minimizers of $\mf$. There are two
main difficulties in doing this. First, the harmonic energy $\mf'$ is not
defined for the functions that we are considering if the domain of integration
includes the axis. Second, we are not dealing with a Dirichlet problem. To
overcome these difficulties an appropriate compact domain is chosen which do
not contain the singularities.  Then a partition function is used to
interpolate between extreme Kerr initial data outside this domain and general
data inside, constructing also an auxiliary intermediate region. This solves
the two difficulties in the sense that now the Dirichlet problem on compact
region can be considered, and the harmonic energy is well defined for this
domain of integration. This allows us to show that the mass functional for Kerr
data is less than or equal to the mass functional for the auxiliary
interpolating data. The final step is to show that as we increase the compact
domain to cover all $\Rt$ the auxiliary data converges to the mass functional
for the original general data. This is the subtle and technical part of the proof.

Finally, we mention that it is possible to construct a heat flow for equations
(\ref{eqq:ha1})--(\ref{eqq:ha2}). Note that the first existence result for
harmonic maps used a heat flow \cite{eells64}.  In our present setting, an
appropriate heat flow that incorporates the singular boundary conditions is
constructed as follows. Consider functions $(\sigma,\omega)$ which depend on an
extra parameter $t$.  Then, we define the following flow
\begin{align}
  \label{eq:haf1}
\dot \sigma  &=  \Delta \sigma  +\frac{e^{-2\sigma}|\partial \omega |^2}{\rho ^4},\\
\label{eq:haf2}
\dot \omega  & =  \Delta \omega   -
2 \frac{\partial \omega\partial \eta}{\eta}.
\end{align}
where a dot denotes partial derivative with respect to $t$.  Equations
(\ref{eq:haf1})--(\ref{eq:haf2}) represent the gradient flow of the energy
(\ref{eqq:5c}).  The important property of the flow is that the energy $\mf$ is
monotonic under appropriate boundary conditions.  This flow have been used in
\cite{Dain:2009qb} as an efficient method for computing numerically both the
solution and the value of the energy $\mf$ at a stationary solution.

\subsection{Two surfaces and the mass functional}
\label{sec:two-surfaces-mass}
In this section we want to define a mass functional over a two-surface
$\Su$. This functional plays a major role in the proofs of the quasi-local
inequalities. Let us motivate first the definition.

Consider the mass functional (\ref{eqq:5c}) defined over $\Rt$. Assume that on
$\Rt$ we have a foliation of two-surfaces. For example, take 
spherical coordinates $(r,\theta,\varphi)$ and the two-surfaces $r=constant$. We can
split the integral (\ref{eqq:5c}) into an integral over the two surface and a
radial integral. However, the integral over the two-surface alone will not have
any intrinsic meaning since, of course, its integrand depends on $r$. To obtain
an intrinsic expression we need somehow to avoid the $r$ dependence taking some
kind of limit. For extreme Kerr initial data this limit is provided naturally
on the cylindrical end. The functions $\sigma$ and $\omega$ have a well defined
and non-trivial limit there (in contrast with the asymptotic flat end where
they tend to zero).  Also, all the radial derivatives go to zero at the
cylindrical end. This surface define an extreme Kerr throat surface (see the
discussion in \cite{Dain:2010uh} and \cite{dain10d} for more details) and it
characterized by only one parameter: the angular momentum
$J$. Explicitly  the functions in that limit are given by
\begin{equation}\label{datos}
 \sigma_0=\ln(4|J|)-\ln(1+\cos^2\theta),\quad
 \omega_0=-\frac{8J\cos\theta}{1+\cos^2\theta}  
\end{equation}
And the area of this two-surface is given by 
\begin{equation}
 A_0=8\pi|J|.
\end{equation}
Consider the following functional over a two-sphere
\begin{equation}
  \label{eqq:22}
 \mfs= \int_0^\pi\left( |\partial_\theta \sigma|^2 +4\sigma + \frac{|\partial_\theta
    \omega|^2}{\eta^2}\right) \sin\theta \, d\theta.
\end{equation}
We can write this integral as an integral over the unit sphere $\mathbb{S}^2$ in
the following way
\begin{equation}
  \label{eqq:44}
  \mfs=\frac{1}{2\pi} \int_{S^2}\left( |D \sigma|^2 +4\sigma + \frac{|D
      \omega|^2}{\eta^2}\right)  \, \ds_0,  
\end{equation}
were $\ds_0=\sin\theta\, d\theta d\phi$ is the area element of the standard
metric in  $\mathbb{S}^2$ and $D$ is the covariant derivative with respect to this
metric. In complete analogy with the discussion in section
\ref{sec:harmonic-maps}, we consider this integral for general functions
$\sigma(\theta, \varphi)$ and $\omega(\theta, \varphi)$  on $\mathbb{S}^2$  which are not
necessarily axially symmetric.  

The Euler-Lagrange equations of this functional are given by
\begin{align}
  \label{eq:18c}
  \Delta_0\sigma-2 & =-\frac{|\partial_\theta \omega|^2}{\tilde\eta^2},\\
  \label{eq:20d}
  \Delta_0\omega  & =2\frac{\partial_\theta\omega\partial_\theta\tilde \eta}{\tilde \eta},
\end{align} 
where
\begin{equation}
  \label{eq:85}
 \tilde \eta=\sin^2\theta \, e^{\sigma}.
\end{equation}

The functional (\ref{eqq:44}) is relevant because the extreme Kerr throat
surface (\ref{datos}) is a solution of the Euler-Lagrange equations.  Equations
(\ref{eq:18c})--(\ref{eq:20d}) can be also deduced from the stationary axially
symmetric equations (\ref{eqq:ha1})--(\ref{eqq:ha2}) taking the limit to the
cylindrical end (given by $r\to 0$ in these coordinates) and using the fall off
behavior of the functions at the cylindrical end, namely
\begin{equation}
  \label{eq:86}
  \sigma =-2\log(r)+ \tilde\sigma(\theta)+O(r^{-1}),\quad \omega=\tilde
  \omega(\theta) +O(r^{-1}).
\end{equation}
Note that the $2$ that appears in the second term in the left hand side of
(\ref{eq:18c}) arises from the characteristic $-2\log(r)$ fall off behavior at the
cylindrical end (we have already discuss this property in equation
(\ref{eq:75})). Also this  $2$  produces the linear term $4\sigma$ in the mass
functional (\ref{eqq:44}).

The value of the functional (\ref{eqq:44}) at the extreme Kerr throat sphere is
given by 
\begin{equation}
 \mfs=8(\ln(2|J|)+1).
\end{equation}

The connection between the mass functional (\ref{eqq:44}) and the energy of
harmonic maps between $\mathbb{S}^2$ and $\mathbb{H}^2$ is very similar as the
one described in the previous section for the mass functional $\mf$.  Namely,
consider the functional
\begin{equation}
  \label{eqq:47}
 \mf'^{\Su}_\Omega=  \frac{1}{2\pi} \int_\Omega \frac{|D \eta|^2+|D
  \omega|^2}{\eta^2}   \, \ds_0, 
\end{equation}
defined on some domain $\Omega\subset  \mathbb{S}^2$, such that $\Omega$ does not include
the poles. Integrating by parts and using  the
identity
\begin{equation}
  \label{eqq:48}
\Ls(\log(\sin\theta))=-1,  
\end{equation}
where $\Ls$ is the Laplacian on $\mathbb{S}^2$, we obtain the following
relation between $\mf$ and $\mf'$
\begin{equation}
  \label{eqq:49}
  \mf'^{\Su}_\Omega= \mf^{\Su}_\Omega+4 \int_\Omega \log\sin\theta\, \ds_0+
\oint_{\partial
    \Omega} (4\sigma + \log\sin\theta) \frac{\partial \log\sin\theta}{\partial
    s}\, dl,
\end{equation}
where $s$ denotes the exterior normal to $\Omega$, $dl$ is the line element
on the boundary $\partial \Omega$ and we have used the obvious notation
$\mfs_\Omega$ to denote the mass functional \eqref{eqq:44} defined over the
domain $\Omega$. The difference between $\mfs$ and $\mf'^{\Su}$ are the boundary
integral plus the second term which is just a numerical constant. Note that if
we integrate over $\mathbb{S}^2  $ this constant term  is finite
\begin{equation}
  \label{eqq:50}
   \int_\Omega \log\sin\theta\, \ds_0=2\log2-2.
\end{equation}
The boundary terms however diverges at the poles. 

In an analogous way as it was described in the previous section, the functional
$\mf'^{\Su}$ defines an energy for maps $(\eta,\omega): \mathbb{S}^2 \to
\mathbb{H}^2$ where $\mathbb{H}^2$ denotes the hyperbolic plane $\{(\tilde
\eta, \omega ) : \tilde \eta > 0\}$, equipped with the negative constant
curvature metric
\begin{equation}
  \label{eqq:51}
  ds^2=\frac{d\tilde\eta^2+d\omega^2}{\tilde\eta^2}. 
\end{equation}
The Euler-Lagrange equations for the energy $\mf'$ are called harmonic maps
from $S^2\to \mathbb{H}^2$. Since $\mfs$ and $\mf'^{\Su}$ differ only by a constant
and boundary terms, they have the same Euler-Lagrange equations.

The variational problem for the mass functional on the two-surface is very
similar to the one for the mass functional on $\Rt$ described in section
\ref{sec:harmonic-maps} (see \cite{Acena:2010ws} for the details).

\section{Open Problems}
\label{sec:open-problems}
In this final section I would like to present the main open problems regarding
these geometrical inequalities. In the light of the recent results of
\cite{Hollands:2011sy} there exists now a very interesting open door to higher
dimensions, but this lies out of the scope of the present review and hence in
this section I will restrict myself to four spacetime dimensions. My aim is to
present open problems which are relevant (and probably involve the discovery of
new techniques) and at the same time they appear feasible to solve.

We begin with the global inequality (\ref{eq:42}). The two main open problems
are the following. 
\begin{itemize}
\item Remove the maximal condition.

\item Generalization for asymptotic flat manifolds with  multiple ends.
\end{itemize}

The situation for the maximal condition in theorems \ref{t:main-1} and \ref{t2}
resembles the strategy of the proof of positive mass theorem by Schoen and Yau
\cite{Schoen79b} \cite{Schoen81} \cite{Schoen81c} \footnote{I thank Marcus
  Khuri for pointing this out to me and for relevant discussion on this
  subject}. That proof was performed first for maximal initial data and then
extended for general data. It is conceivable that similar techniques (i.e. the
use of Jang equation) can be used here also, but it is far from obvious how to
extend these ideas to the present case.

The most relevant open problem regarding the global inequality (\ref{eq:42}) is
its validity for manifolds with multiple asymptotic ends with non-trivial
angular momentum.  The physical heuristic argument presented in section
\ref{sec:physical-picture} applies to that case and hence there little doubt
that the inequality holds.  In particular, as we already mentioned in section
\ref{sec:global-inequality}, for the case of three asymptotic ends there are
strong numerical evidences for the validity of the inequality
\cite{Dain:2009qb}.

Theorem \ref{Piotr-Gilbert} (proved in \cite{Chrusciel:2007ak}) reduces the
proof of the inequality to prove the bound (\ref{eq:49}) for the mass
functional evaluated at the stationary solution.  The case of three asymptotic
ends (which, roughly speaking, is equivalent to say that we have two black
holes) is special for the following reason. There exists exact stationary
axially symmetric solutions of the Einstein equations (the so-called
double-Kerr-NUT solutions \cite{kramer80b} \cite{neugebauer80}) which represent
two Kerr-like black holes. These solutions contain singularities, that prevent
them to qualify as genuinely equilibrium state for binary black holes.  In
fact, one of the main part in the strategy to prove that the uniqueness theorem
hold for the binary case is to prove that these solutions are always singular
\cite{Neugebauer:2011qb}. However, even if these solutions are singular they
can be useful to prove the bound (\ref{eq:49}), because in order to qualify as
a stationary point of the mass functional $\mf$ all we need is that the
functions $(\eta, \omega)$ are regular. It is conceivable that some of these
exact solutions have this property (that means, of course, that other
coefficient of the metric are singular) and hence with the explicit expression
for $(\eta, \omega)$ provided by them it will possible to evaluate $\mf$ and
check the bound (\ref{eq:49}). In the articles \cite{Manko:2011qh}
\cite{CabreraMunguia:2010uu} the geometrical inequality (\ref{eq:42}) has been
studied for these exact solutions.  These results provide a guide for which
subclass of these solutions are potentially useful to prove the bound
(\ref{eq:49}). Unfortunately the solutions, although explicit, are very
complicated and it is very difficult to compute the mass functional $\mf$ for
them\footnote{I thank Piotr Chrusciel for relevant discussion on this point}.

To compute the value of $\mf$ for this kind of exact solution would be
certainly a very interesting result which not only will prove the inequality
(\ref{eq:42}) for the three asymptotic ends case but also will hopefully
provide some new interpretation of the double-Kerr-NUT solutions. However this
result will be confined to the three asymptotic ends case and probably will not
yield light into the mechanism of the variational problem for the mass
functional $\mf$ with multiple ends.  The basic property which is expected to
satisfies $\mf$ is the following: if an extra black hole is added, with arbitrary
angular momentum, then the value of $\mf$ increase.  This property is of course
another way of saying that the force between the black holes is always
attractive (in particular, it can not be balanced by spin-spin repulsive
interaction).  The variational problem for the mass functional $\mf$ with
multiple ends appears to have a remarkably structure. In particular, there is
formal similarity between this problem and the kind of singular boundary
problems for harmonic maps studied in \cite{bethuel94}.

We mention in section \ref{sec:physical-picture} that there is a clear physical
connection between the global inequality (\ref{eq:42}) and the Penrose
inequality with angular momentum in axial symmetry. Hence, it appropriate to
list here the Penrose inequality as a relevant open problem for axially
symmetric black holes (for more detail on this problem see the review
\cite{Mars:2009cj}):
\begin{itemize}
\item Prove the Penrose inequality with angular momentum Eq. (\ref{eq:23}). 
\end{itemize}
However, it is important to emphasize that it not clear that the techniques
used to prove theorems \ref{t:main-1} and \ref{t2} will help to solve this
problem. The reason is the following.  The proof of the Penrose inequality
involve an inner boundary, namely the black hole horizon.  On the other hand,
theorems \ref{t:main-1} and \ref{t2} refer to complete manifolds without inner
boundaries. As we mention in \ref{sec:global-inequality} there are results in
axial symmetry which includes inner boundaries (i.e. \cite{Gibbons06}
\cite{Chrusciel:2011eu}) and use similar techniques as in theorems
\ref{t:main-1} and \ref{t2} (namely, the representation of the mass as positive
definite integral in axial symmetry, see section
\ref{sec:mass-axial-symmetry}). However the boundary for the Penrose inequality
has a very important property: it should be an outer minimal surface (for
simplicity, we discuss only the Riemannian case). This property is very
difficult to incorporate in an standard boundary value problem.  To see if the
mass formula in axial symmetry is useful to prove the Penrose inequality with
angular momentum the natural first step is to prove the Riemannian Penrose
inequality without angular momentum in axial symmetry using this mass
formula. The results presented in \cite{Gibbons06} \cite{Chrusciel:2011eu}
contribute in this direction, but so far the problem remains open.  May be
there exists a combination of the global flows techniques developed in
\cite{Huisken01} \cite{Bray01} for the Riemannian Penrose inequality with the
mass functional that incorporate the angular momentum in axial symmetry. For
example, it is suggestive that the strategy for the proof of the charged
Penrose inequality Eq.  (\ref{eq:22}) given in \cite{Jang79} and
\cite{Huisken01} consists in first prove the inequality (\ref{eq:53}) using a
flow and a lower bound to the scalar curvature that resemble the mass
functional. However, a generalization of this construction to include the
angular momentum is far from obvious.

We turn now to the quasi-local inequalities. The three main problems are the following. 

\begin{itemize}
\item Include the charge and the electromagnetic angular momentum in axial
  symmetry for the inequality (\ref{e:inequality}). 

\item Isoperimetric inequalities in axial symmetry with angular momentum. That
  is, a version for theorem  \ref{t:main3}  (in axial symmetry) with
  angular momentum instead of charge.

\item A generalization of the inequality (\ref{e:inequality}) without axial
  symmetry. 

\end{itemize}

The inclusion of charge in the inequality (\ref{e:inequality}) is important, of
course, since charge is the other relevant parameter that characterized the
Kerr-Newman black hole.  But is also relevant for another reason. The angular
momentum that appears in this inequality is the Komar gravitational angular
momentum. In the generalization with charge it is expected that the total
angular momentum (i.e. gravitational plus electromagnetic) appear in this
inequality.  This is shown, including also the magnetic charge, in
\cite{gabach11}.  In addition, this is important in connection with the
rigidity statement.  There is work in progress on this problem \cite{GabJar11}.

We mention in section \ref{sec:physical-picture} that a version of the
inequality (\ref{e:inequality}) for isoperimetric surfaces (instead of black
hole horizons) could have interesting astrophysical applications, since
apparently neutron stars are close to saturate this kind of inequalities. Such
theorem will be analogous to theorem \ref{t:main3} for the charge.  However, it
is by no means clear that similar techniques as the one used in the proofs of
theorem \ref{t:main-2} can be applied to that case.

Finally we mention the problem of finding versions of inequality
(\ref{e:inequality}) without any symmetry assumption. In contrast with the
other open problems presented here, this is not a well defined mathematical
problem since there is no unique notion of quasi-local angular momentum in the
general case.  However to explore the scope of the inequality in regions close
to axial symmetry (in some appropriate sense) can perhaps provide such a
notion. From the physical point of view  I do not see any reason why this
inequality should only hold in axial symmetry.

\section*{Acknowledgments}
I would like to thank Robert Geroch, Jos\'e Luis Jaramillo, Carlos Kozameh and
Walter Simon for illuminating discussions during the preparation of this
review. 

The author is supported by CONICET (Argentina). This work was supported in part
by grant PIP 6354/05 of CONICET (Argentina), grant Secyt-UNC (Argentina) and
the Partner Group grant of the Max Planck Institute for Gravitational Physics
(Germany).


\end{document}